\documentclass[twocolumn,prb,color,superscriptaddress,psfig,showpacs,amsmath,amssymb,fleqn,nobibnotes]{revtex4-1}

\setcitestyle{square,numbers}

\usepackage{graphicx}% Include figure files
\usepackage[caption=false]{subfig}
\usepackage{float}
\usepackage{epstopdf}
\usepackage{dcolumn}% Align table columns on decimal point
\usepackage{hyperref}% add hypertext capabilities; optional: [hidelinks]
\hypersetup{
    colorlinks   = true, %Colours links instead of boxes
    urlcolor     = blue, %Colour for external hyperlinks
    linkcolor    = blue, %Colour of internal links
    citecolor    = blue %Colour of citations
}
\usepackage{xcolor}
\usepackage{soul}
\setstcolor{red}
\usepackage{natbib}
\usepackage{amsmath}

\newcommand{\CGT}{Cr$_2$Ge$_2$Te$_6$}
\newcommand{\gc}{$^\circ$C}

\begin{document}
	
\title{Magnetic anisotropy and spin-polarized two-dimensional electron gas\\ in the van der Waals ferromagnet \CGT}

\author{J.~Zeisner}
\thanks{These authors contributed equally to this work.}
\affiliation{Leibniz Institute for Solid State and Materials Research IFW Dresden, 01069 Dresden, Germany}
\affiliation{Institute for Solid State and Materials Physics, TU Dresden, 01062 Dresden, Germany}
\author{A.~Alfonsov}
\thanks{These authors contributed equally to this work.}
\affiliation{Leibniz Institute for Solid State and Materials Research IFW Dresden, 01069 Dresden, Germany}
\author{S.~Selter}
\affiliation{Leibniz Institute for Solid State and Materials Research IFW Dresden, 01069 Dresden, Germany}
\affiliation{Institute for Solid State and Materials Physics, TU Dresden, 01062 Dresden, Germany}
\author{S.~Aswartham}
\affiliation{Leibniz Institute for Solid State and Materials Research IFW Dresden, 01069 Dresden, Germany}
\author{M.~P.~Ghimire}
\affiliation{Central Department of Physics, Tribhuvan University, Kirtipur, 44613, Kathmandu, Nepal}
\affiliation{Leibniz Institute for Solid State and Materials Research IFW Dresden, 01069 Dresden, Germany}
\affiliation{Condensed Matter Physics Research Center, Butwal-11, Rupandehi, Lumbini, Nepal}
\author{M.~Richter}
\affiliation{Leibniz Institute for Solid State and Materials Research IFW Dresden, 01069 Dresden, Germany}
\affiliation{Dresden Center for Computational Materials Science (DCMS), TU Dresden, 01062 Dresden, Germany}
\author{J.~van den Brink}
\affiliation{Leibniz Institute for Solid State and Materials Research IFW Dresden, 01069 Dresden, Germany}
\author{B.~B\"{u}chner}
\affiliation{Leibniz Institute for Solid State and Materials Research IFW Dresden, 01069 Dresden, Germany}
\affiliation{Institute for Solid State and Materials Physics, TU Dresden, 01062 Dresden, Germany}
\author{V.~Kataev}
\affiliation{Leibniz Institute for Solid State and Materials Research IFW Dresden, 01069 Dresden, Germany}
\date{\today}

\begin{abstract}

We report a comprehensive experimental investigation on the magnetic anisotropy in bulk single crystals of \CGT, a quasi-two-dimensional ferromagnet belonging to the family of magnetic layered transition metal trichalcogenides that have attracted recently a big deal of interest with regard to the fundamental and applied aspects of two-dimensional magnetism. For this purpose electron spin resonance (ESR) and ferromagnetic resonance (FMR) measurements have been carried out over a wide frequency and temperature range. A gradual change in the angular dependence of the ESR linewidth at temperatures above the ferromagnetic transition temperature $T_{\rm c}$ reveals the development of two-dimensional spin correlations in the vicinity of $T_{\rm c}$ thereby proving the intrinsically low-dimensional character of spin dynamics in \CGT. Angular and frequency dependent measurements in the ferromagnetic phase clearly show an easy-axis type anisotropy of this compound. Furthermore, these experiments are compared with simulations based on a phenomenological approach, which takes into account results of static magnetization measurements as well as high temperature $g$ factors obtained from ESR spectroscopy in the paramagnetic phase. As a result the determined magnetocrystalline anisotropy energy density (MAE) $K_U$ is \mbox{$(0.48 \pm 0.02) \times10^6$\, erg/cm$^3$}. This analysis is complemented by density functional calculations which yield the experimental MAE value for a particular value of the electronic correlation strength $U$. The analysis of the electronic structure reveals that the low-lying conduction band carries almost completely spin-polarized, quasi-homogeneous, two-dimensional states.
	
\end{abstract}

\maketitle

\section{Introduction}

Layered van der Waals crystals constitute a large class of materials which is intensively studied in the field of current solid state research \cite{Xu2013,Novoselov2016}. Weak couplings between individual layers render these quasi-two-dimensional compounds optimal as a starting point for fabrication of two-dimensional (2D) atomic crystals \cite{Novoselov2005} which show a plethora of electronic properties and, consequently, a high potential for technical applications (see, e.g., Refs.~\cite{Xu2013,Novoselov2016,Geim2013} and references therein). Recently, some members of the family of layered transition metal trichalcogenides (TMTC) came into focus of research due to their peculiar magnetic properties in combination with semiconducting or insulating behavior \cite{Sivadas2015, Williams2015,Zhuang2015,Wang2016,Yu2018,Kim2018}. In particular, TMTC are currently considered as promising materials for future technological applications but also  offer an extensive materials base for exploring fundamental magnetic properties of strongly correlated 2D electron systems. \CGT \ belongs to this class of materials and it was first synthesized by V. Carteaux \textit{et al.} \cite{Carteaux1995}. The carriers of magnetism in this material are Cr$^{3+}$ ions with spin $S = 3/2$ which are octahedrally coordinated by Te ligands and form a 2D  honeycomb-like magnetic lattice in the $ab$ plane. In the bulk crystal individual magnetic layers stack along the crystallographic $c$ axis and are only weakly coupled by van der Waals forces.  By virtue of intrinsic magnetic two-dimensionality \CGT\ has been proposed, e.g., as magnetic substrate for nanoelectronic devices \cite{Ji2013} or as a candidate for next generation memory devices \cite{Hatayama2018}. From the fundamental point of view the observation of intrinsic ferromagnetism in atomically thin films of \CGT \ is intriguing \cite{Gong2017}. Such phenomenon was also observed in single layers of the related materials CrI$_3$ \cite{Huang2017}, and VSe$_2$ \cite{Bonilla2018}.

Motivated by the above-mentioned aspects, \CGT \ received considerable attention both experimentally \cite{Carteaux1995,Zhang2016,Tian2016,Lin2017,Liu2017,Lin2018,Sun2018,Liu2018} and theoretically \cite{Li2014,Sivadas2015} during the last years. Aiming at a detailed understanding of the magnetic properties of this material, in bulk as well as in monolayer form, a quantitative characterization of the magnetic anisotropies represents a key task. Regarding the latter, first insights were recently obtained by X. Zhang \textit{et al.} \cite{Zhang2016} for the case of bulk samples by studying ferromagnetic resonance (FMR) at frequencies below 20\,GHz as a function of temperature with external magnetic field applied in the $ab$ plane. In the present work we report a comprehensive investigation of the magnetic anisotropy of bulk \CGT \ single crystals by means of electron spin resonance (ESR) spectroscopy over a wide frequency and temperature range for several field orientations which allows a refined quantitative analysis of the magnetocrystalline anisotropy energy density (MAE). Furthermore, we compare experimentally obtained MAE with our results from density functional (DF) calculations and find a good agreement for a particular strength of electronic correlations $U$. In addition, the two-dimensionality of spin dynamics in \CGT \ in the vicinity of the ferromagnetic transition temperature $T_{\rm c}$ is proven by angular dependent measurements of the ESR linewidth at various temperatures.

Previous theoretical \cite{Li2014,Wang2018} and experimental \cite{Wang2018} work suggests strong spin polarization of the electronic states of \CGT \ at the valence band maximum (VBM) and the conduction band minimum (CBM). Since these investigations propose different relative polarizations at VBM and CBM, we use the obtained electronic structure results for a further analysis.

The paper is organized as follows. Details of synthesis and characterization of the samples as well as further experimental and calculation details are provided in Sec.~\ref{sec:methods}. Experimental and theoretical findings are presented and discussed in Sec.~\ref{sec:results_and_discussion}. Main conclusions of this study are drawn in Sec.~\ref{sec:conclusions}. Appendices \ref{sec:sample_details} and \ref{sec:add_x-band} contain additional information on the samples used for the measurements and further results obtained from low-frequency FMR, respectively.

\section{Samples and Methods}
\label{sec:methods}

Single crystals of \CGT \ were grown by using the self flux technique described by X. Zhang \textit{et al.} \cite{Zhang2016}. A mixture of chromium granules (4N, MaTeck), germanium chips (5N, Sigma Aldrich) and tellurium lumps (5N, Alfa Aesar) with molar ratios of Cr\,:\,Ge:Te\,=\,10\,:\,13\,:\,77 were taken in an alumina crucible which was then sealed in a quartz ampule under a partial atmosphere of Ar (approx. 300\,mbar). Then, the quartz ampule was placed upright in a Muffel furnace (Nabertherm) and heated to 1000\,\gc \ for 24\,h followed by cooling with a rate of 2\,\gc/h to 450\,\gc. At this temperature excessive melt was centrifuged. Shiny, plate-like crystals of up to 11\,mm $\times$ 10\,mm $\times$ 0.2\,mm were obtained.

\begin{figure}[!tbp]
	\includegraphics[width=\columnwidth]{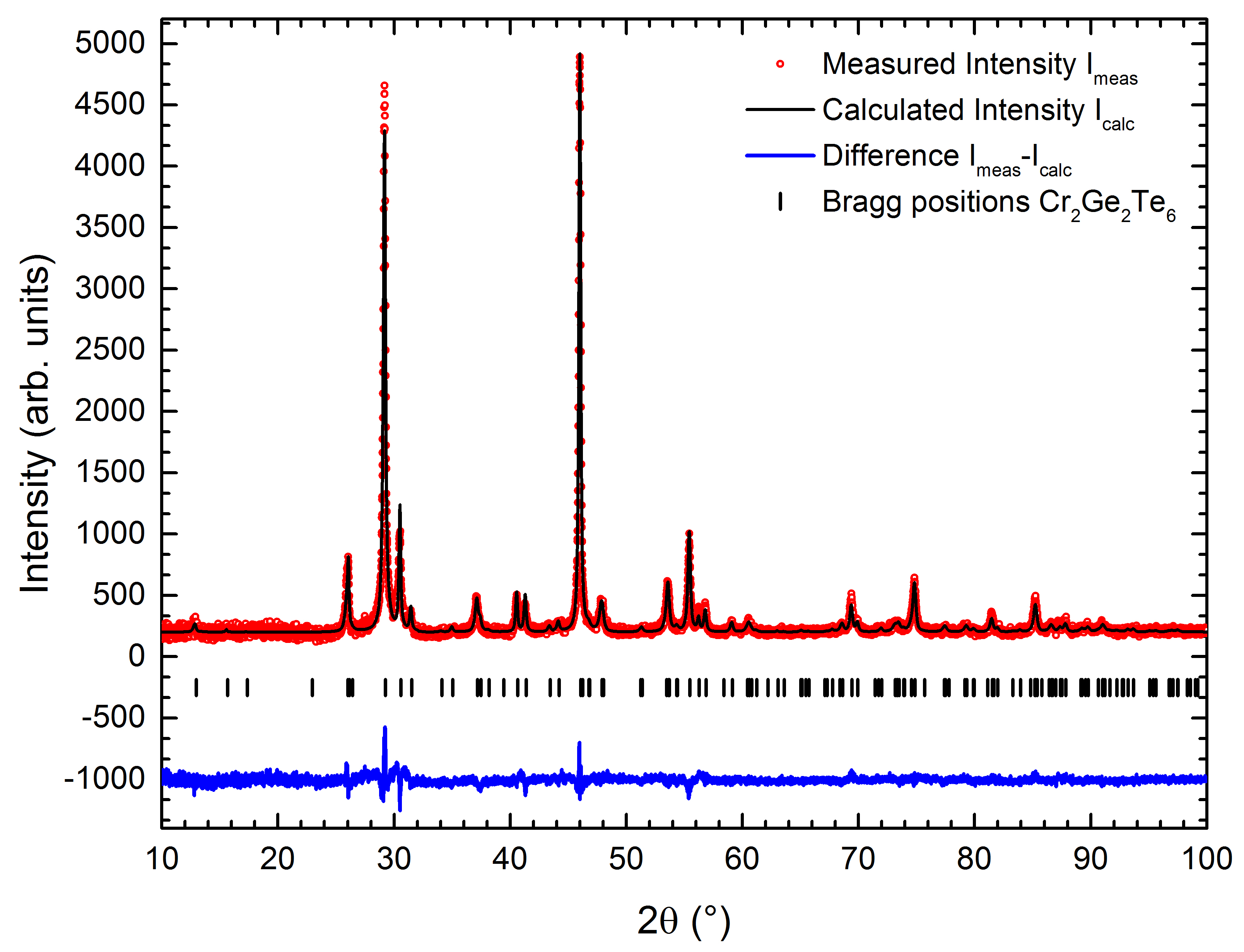}
	\caption{Powder X-ray diffraction pattern from Cu-K$_{\alpha_1}$ radiation (1.54059 \r{A}) of pulverized crystals of \CGT\ (red dots), calculated pattern from Rietveld analysis (black line), difference between measured and calculated intensity (blue line) and expected Bragg positions for the \CGT\ structure (black bars).}
	\label{fig:CGT_XRD_Rietveld}
\end{figure}

Crystals were characterized by powder X-ray diffraction (PXRD) and energy dispersive X-ray spectroscopy (EDX). EDX was performed at an accelerating voltage of 30\,kV using a ZEISS EVO MA 10 scanning electron microscope (SEM) with an energy dispersive X-ray analyzer. A mean elemental composition of Cr\,:\,Ge\,:\,Te\,=\,(20.9\,$\pm$0.2)\,\%\,:\,(20.4\,$\pm$0.2)\,\%\,:\,(58.7\,$\pm$0.1)\,\% was obtained by EDX. This is in good agreement with the expected elemental composition of Cr\,:\,Ge\,:\,Te\,=\,20\,\%\,:\,20\,\%\,:\,60\,\% assuming a general uncertainty of the method of approximately $\pm$2\% for each element. In the range of this uncertainty, the low standard deviation of the chemical composition over all measured spots indicates a homogeneous elemental distribution.
PXRD was performed on pulverized crystals on a STOE STADI laboratory diffractometer in transmission geometry with Cu K$_{\alpha_1}$ radiation from a curved Ge(111) single crystal monochromator and detected by a 6$^\circ$-linear position sensitive detector manufactured by STOE \& Cie. Rietveld analysis was carried out on the measured X-ray pattern using the Jana2006 software package \cite{Petricek2014}, the diffraction pattern with the fitted model is shown in Fig.~\ref{fig:CGT_XRD_Rietveld}. Structural parameters and corresponding R-values are summarized in Table~\ref{tab:Rietveld1} in Appendix~\ref{sec:sample_details}. These results are in good agreement with literature \cite{Carteaux1995,Yang2016}.

DC magnetization was measured as a function of temperature and magnetic field using a vibrational sample quantum interference device magnetometer (SQUID-VSM) from Quantum Design. All measurements were carried out with a vibrational amplitude of 2\,mm. The values obtained for magnetic moments were corrected due to deviation of the measured sample shape and size from a point dipole. This correction follows the procedure described in Ref.~\cite{VSM_corr} and further details are presented in Appendix~\ref{sec:sample_details}. For temperature dependent measurements, the sample was demagnetized at 300\,K by applying an external field of 7\,T and reducing it to zero while periodically switching the polarity of the field. After that the sample was zero field cooled (ZFC) down to 1.8\,K. The measurement itself was performed upon warming from 1.8\,K to 300\,K with an external applied field of 0.35\,T or 3\,T, respectively. For field sweep experiments the sample was demagnetized as described above and cooled to the desired temperature of 4\,K or 7\,K, respectively. During the measurement the field was ramped from zero to 7\,T to obtain a virgin magnetization curve, followed by a measurement of the full hysteresis curve by cycling the field from  7\,T to -7\,T and back.

\begin{table*}[tbh]
	\caption{Comparison of atomic positions of \CGT \ as \\determined from PXRD and DF calculations.}
	\setlength\extrarowheight{0.5pt}
	\begin{ruledtabular}
		\begin{tabular}{c c c c c c}
			Method & Atom & Wyckoff site & $x$ & $y$ & $z$\\
			\hline
			& Te & 18$f$ & -0.0017(11) & 0.3642(6) & 0.0853(10) \\
			PXRD & Cr & 6$c$ & 0 & 0 & 0.3330(30) \\
			& Ge & 6$c$ & 0 & 0 & 0.0588(14) \\
			\hline
			& Te & 18$f$ & -0.003 & 0.374 & 0.085\\
			GGA & Cr & 6$c$ & 0 & 0 &  0.334 \\
			& Ge & 6$c$ & 0 & 0 & 0.059 \\
			\hline
			& Te & 18$f$ & -0.003 & 0.373 & 0.085 \\
			GGA+U,& Cr & 6$c$ & 0 & 0 & 0.334 \\
			$U$ = 2\,eV& Ge& 6$c$ & 0 & 0 & 0.058 \\
		\end{tabular}
	\end{ruledtabular}
	\label{tab:comparison_atomic_positions}
\end{table*}

ESR and FMR experiments were performed using three different setups. For high-field/high-frequency ESR (HF-ESR) measurements a home made setup comprising a superconducting magnet equipped with a variable temperature insert (Oxford Instruments), a vector network analyzer (PNA-X from Keysight Technologies) for generation and detection of microwaves, and oversized waveguides were used. This setup allows measurements over a wide frequency (20 - 330\,GHz), magnetic field (0 - 16\,T), and temperature (1.8 - 300\,K) range. All HF-ESR measurements were performed in transmission mode using a Faraday configuration. Resonance fields were determined from the position of the minimum in the transmission. In addition, a cantilever-based torque detected ESR setup was employed. In this setup magnetic fields up to 10\,T and temperatures between 7 and 300\,K are provided by a magneto-optical cryostat (Oxford Instruments). Microwaves with frequencies of \mbox{$\sim$ 63\,GHz} and \mbox{$\sim$ 125\,GHz} are generated by an amplifier/multiplier chain (AMC from Virginia Diodes Inc.) and are focused on the sample by means of the microwave optics. The sample is placed on the tip of a piezoresistive cantilever (see Appendix~\ref{sec:sample_details}), whose vibration amplitude and frequency changes are used to detect the FMR signal. Furthermore, a standard X-band ESR-spectrometer (EMX from Bruker) operating at a microwave frequency of 9.6\,GHz and providing magnetic fields up to 0.9\,T was used. It is equipped with a He-gas flow cryostat (Oxford Instruments) and a goniometer allowing temperature and angular dependent measurements between 4 and 300\,K.

The DF calculations were carried out with the all-electron full-potential local-orbital (FPLO) code \cite{Koepernik1999,fplo}, version 18.00, using
the standard generalized-gradient approximation (GGA) in the parameterization of Perdew, Burke, and Ernzerhof (PBE-96) \cite{Perdew1996}.
The GGA+$U$ functional with atomic-limit (AL) double-counting corrections \cite{Czyzyk1994} was applied to the Cr-$3d$ orbitals in a part of the calculations.
The AL flavor of double-counting is in general to be preferred against the around-mean-field version since it fulfills Hund's first rule \cite{Ylvisaker2009}.
The values of $U$ were varied in the range from 0.5\,eV to 4.0\,eV, while the related exchange parameter $J$ was fixed to 0.6\,eV.
All $k$-space integrations were carried out with the linear tetrahedron method and $12\times 12\times 12$ subdivisions in the full Brillouin zone. The densities of states were evaluated with a refined mesh of $24\times 24\times 24$ subdivisions. The experimental space group R$\bar{3}$ (no. 148) and lattice parameters as obtained in the present experiments were used for all calculations. Starting from the experimentally determined values, the atomic positions were optimized by scalar-relativistic
GGA or GGA+$U$ calculations for each value of $U$ until all residual forces were smaller than 1\,meV/\AA{}. Atomic positions as obtained from PXRD and two of the DF calculations are compared in Table~\ref{tab:comparison_atomic_positions}. As can be seen from this table, the refined calculated atomic positions agree with their experimental counterparts except for the Te-$y$ coordinates, which differ by 0.01. This difference could be caused by omitting spin-orbit coupling in the optimization. For the determination of magnetic anisotropy energy, self-consistent calculations using the optimized structure were carried out in full relativistic mode for both of the considered orientations of the magnetization, [001] (perpendicular to the quasi-2D plane) and [100] \mbox{(arbitrarily chosen in the plane).}

\section{Results and discussion}
\label{sec:results_and_discussion}
\subsection{Magnetization}
\label{subsec:magnetization}

\begin{figure}[!tb]
	\centering
	\includegraphics[width=0.95\columnwidth]{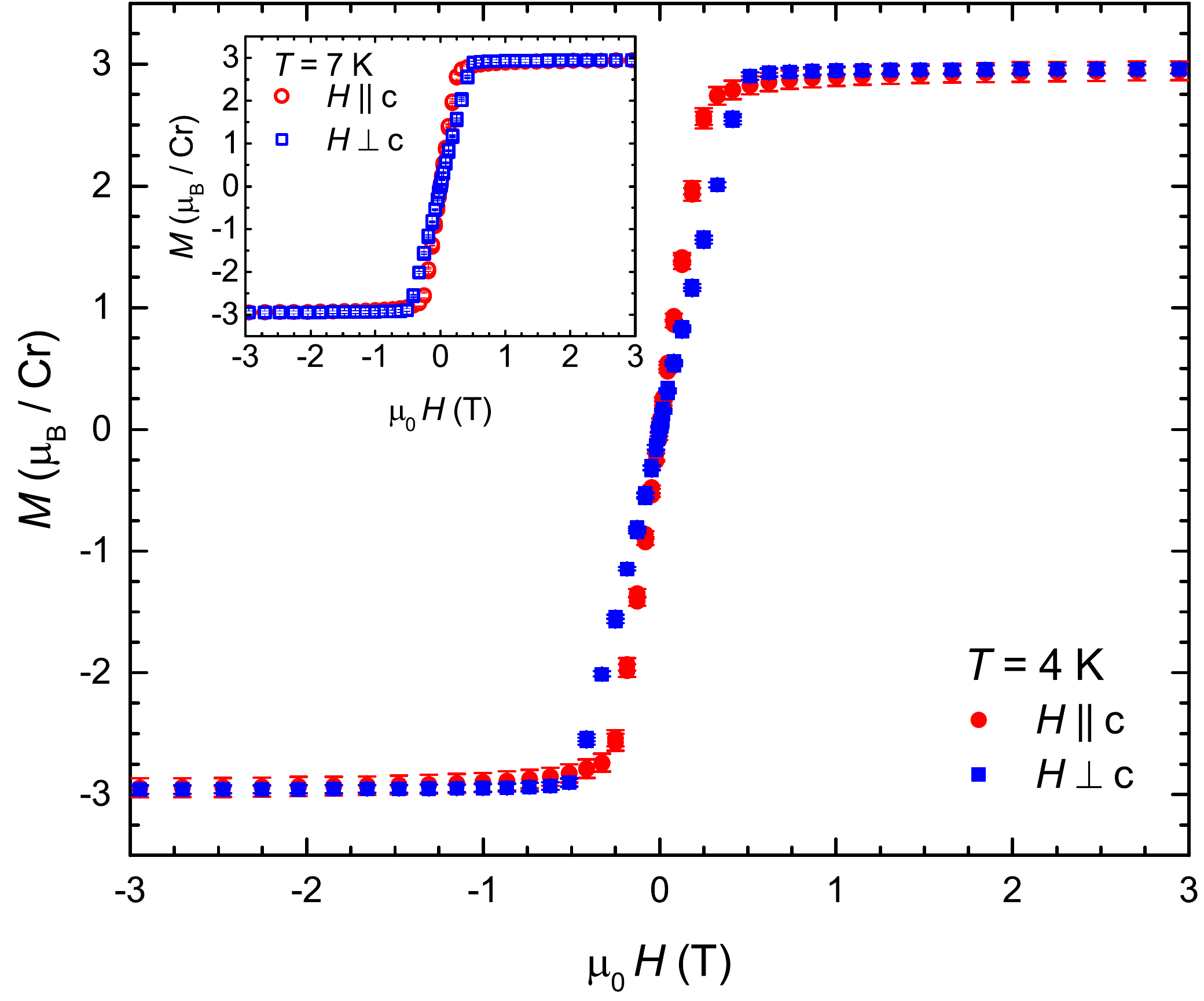}
	\caption{Magnetization as a function of external magnetic field applied parallel (red circles) and perpendicular (blue squares) to the $c$ axis at 4\,K. Corresponding curves recorded at 7\,K are shown in the inset.}
	\label{fig:M_vs_H}
\end{figure}

Magnetization $M$ measured as a function of external magnetic field $H$ is shown in Fig.~\ref{fig:M_vs_H}. Measurements were conducted with magnetic field aligned parallel and perpendicular to the $c$ axis at temperatures of 4 and 7\,K, respectively. At both temperatures a similar behavior is found: While raising the field from zero, $M(H)$ rapidly increases until the onset of saturation at fields $H_{\rm sat}^{\rm ab} \approx  0.5$\,T and $H_{\rm sat}^{\rm c} \approx 0.3$\,T for  $H \perp c$ and  $H\parallel c$ at 4\,K, respectively. The faster saturation of the magnetization for $H\parallel c$ ($H_{\rm sat}^{\rm c} < H_{\rm sat}^{\rm ab}$) evidences the easy-axis type anisotropy of \CGT. Furthermore, all measured magnetization curves reach an isotropic saturation value of about $3 \ \mu_B$ per Cr ion which is in very good agreement with the value theoretically expected for Cr$^{3+}$ ions with $S = 3/2$ and a $g$-factor of~2. 	

\begin{figure}[!tbh]
	\centering
	\includegraphics[width=\columnwidth]{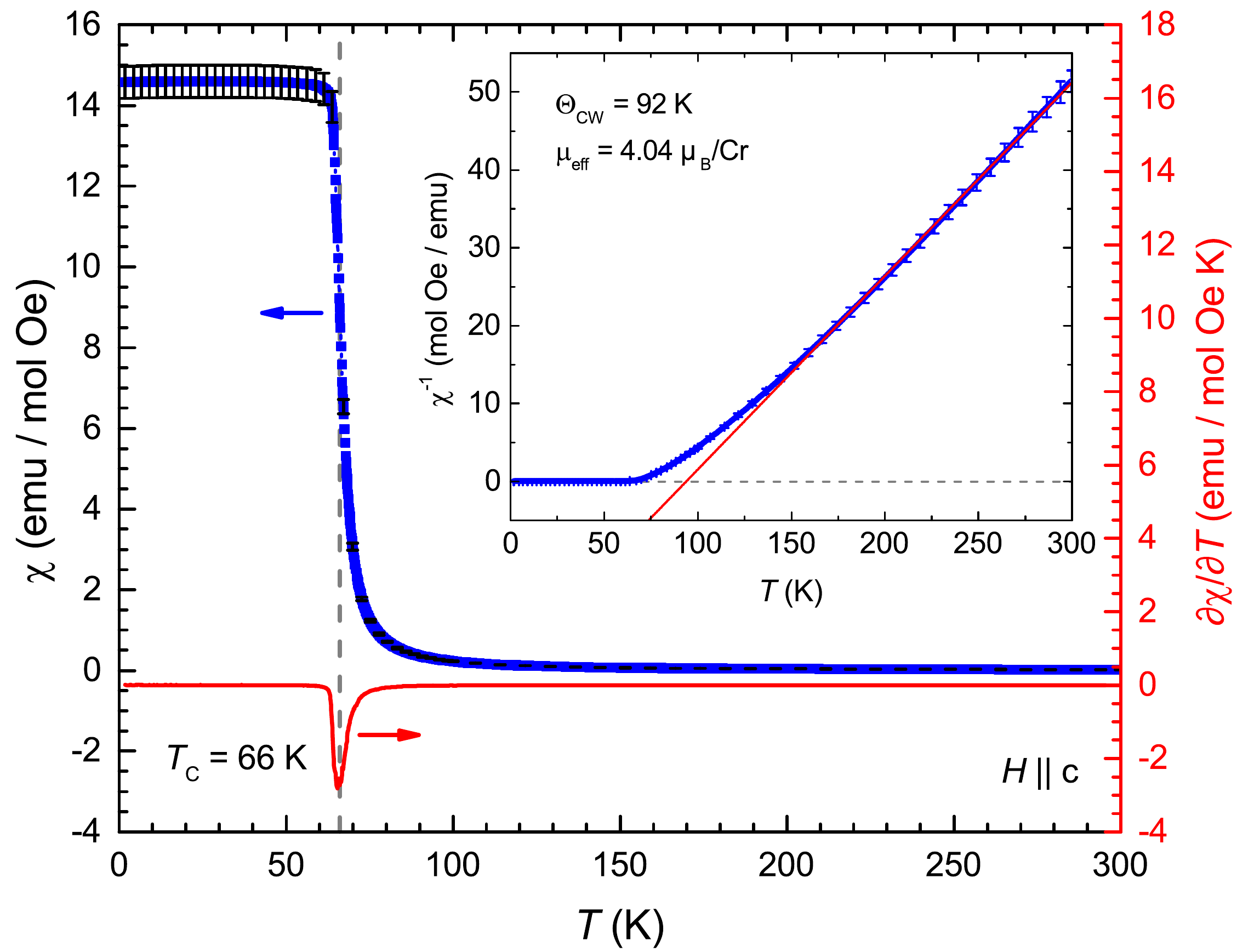}
	\caption{ Temperature dependence of the static magnetic susceptibility measured with an external field of 0.1\,T oriented parallel to the $c$ axis (left axis of the main panel). The right axis represents the first derivative $\partial \chi / \partial T$, which shows a minimum at $T_{\rm c}$. The inverse susceptibility is given together with a linear fit in the inset. For better visibility, error bars of $\chi$ and $1/\chi$ are only shown for every 25$^{\rm th}$ data point.}
	\label{fig:determination_transition_temp}
\end{figure}

The ferromagnetic transition temperature $T_{\rm c}$ was determined from measurements of the temperature-dependent magnetization in an external field of 0.1\,T applied parallel to the $c$ axis from which the static magnetic susceptibility $\chi$ was calculated (Fig.~\ref{fig:determination_transition_temp}). From the position of the minimum in the first derivative of the susceptibility $\partial \chi / \partial T$, a $T_{\rm c}$ of $(66 \pm 1)$\,K was obtained. The uncertainty in $T_{\rm c}$ is caused by a finite width of the minimum, see Fig.~\ref{fig:determination_transition_temp}. Our value of  $T_{\rm c}$ is consistent with the results of Ref.~\cite{Carteaux1995} where, depending on the method of determination, transition temperatures of 61\,K and 65\,K were found. Similar values were reported for $T_{\rm c}$ in other recent studies on \CGT \ \cite{Liu2017,Lin2017,Liu2018}. The inverse susceptibility is shown as a function of temperature in the inset of Fig.~\ref{fig:determination_transition_temp} together with a linear fit according to the Curie-Weiss-law. This fit yields a Curie-Weiss temperature $\Theta_{\rm cw}$ of $(92 \pm 1)$\,K and an effective moment $\mu_{\rm eff}$ of $(4.04 \pm 0.01)$\,$\mu_B$/Cr. The latter is consistent with the value 3.87\,$\mu_B$/Cr theoretically expected for a spin-only system with a $g$-factor of 2. The deviation between $T_{\rm c}$ and $\Theta_{\rm cw}$ is in agreement with previous studies \cite{Lin2017,Liu2017,Ji2013} and can be attributed to the development of short-range correlations already above $T_{\rm c}$ as will be discussed in Sec.~\ref{subsec:results_ESR}.

\begin{figure*}[!htbp]
	\centering
	\includegraphics[width=\textwidth]{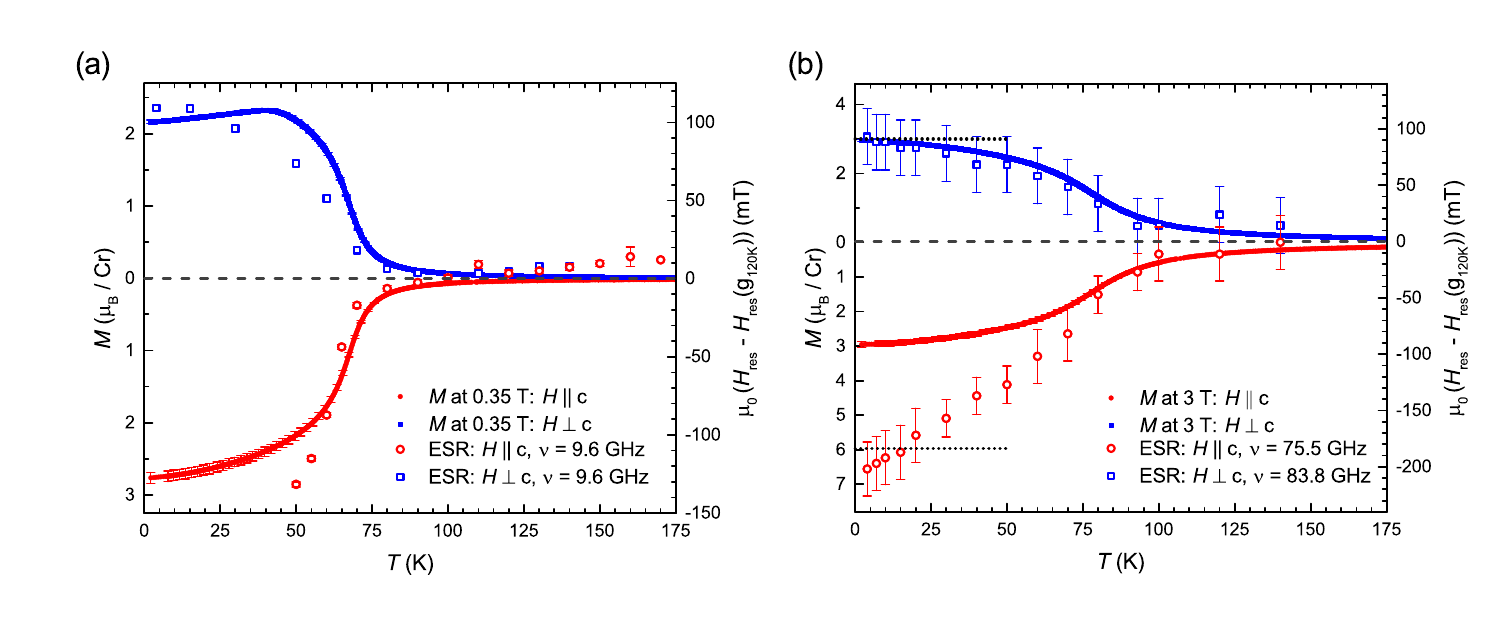}
	\caption{Resonance shift $\delta H$ (right vertical scale) as a function of temperature at a microwave frequency $\nu$ of 9.6\,GHz (a) and at higher frequencies (b) for external magnetic field applied parallel and perpendicular to the $c$ axis, respectively (open symbols) together with the $T$-dependence of the static magnetization $M(T)$ (left vertical scale) for the respective field orientations (closed symbols). For better visibility, error bars for static magnetization are shown only for every tenth data point. The zero-shift of an ideal paramagnet is indicated by a gray dashed line. The black dotted line on panel (b) denotes the shift at 4\,K obtained from the analysis of the frequency dependent FMR data (see the text).}
	\label{fig:T-dep}
\end{figure*}

Furthermore, results of temperature dependent measurements in different magnetic fields are presented in Fig.~\ref{fig:T-dep} for two orientations of the field with respect to the crystal axes. At an external field of 0.35\,T, i.e., below the saturation field in the $ab$ plane at low temperatures, an anisotropic behavior of the magnetization is found. For magnetic fields applied perpendicular to the $c$ axis a maximum of the magnetization is observed around 40\,K, whereas magnetization increases monotonously with decreasing temperature for $H \parallel c$. A similar difference between measurements with fields perpendicular and parallel to the easy direction was found in other Cr-based 2D honeycomb materials, for example CrCl$_3$ \cite{McGuire2017} and CrI$_3$ \cite{McGuire2015,Richter2018}. This behavior might be caused either by antiferromagnetic couplings between the ferromagnetic layers, a more complex ferromagnetic state in low magnetic fields or a temperature dependence of the magnetic anisotropy \cite{McGuire2017,McGuire2015,Richter2018}. On the other hand, for a field of 3\,T, which is larger than the low-temperature saturation fields for both orientations in \CGT, an isotropic behavior was found with the onset of ferromagnetic (FM) correlations being shifted to higher temperatures as the external magnetic field enhances FM correlations.

\subsection{ESR and FMR}
\label{subsec:results_ESR}
 
\begin{figure*}[!htbp]
	\centering
	\includegraphics[width=\textwidth]{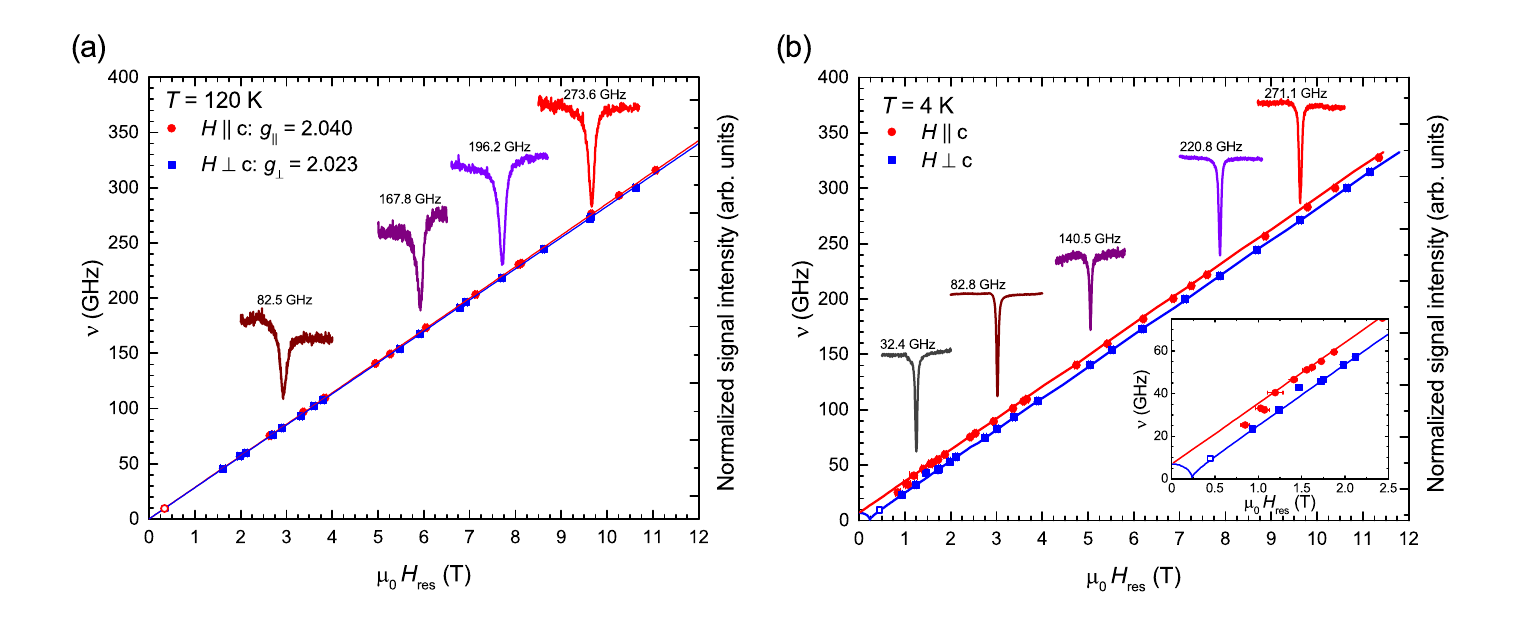}
	\caption{Relation between the excitation frequency $\nu$ (left vertical scale) and the corresponding resonance field $\mu_0H_{\text{res}}$ at 120\,K (a) and 4\,K~(b) (closed symbols) for external magnetic field applied parallel and perpendicular to the $c$ axis. Representative field-sweep ESR spectra are shown for various $\nu$ (right vertical axis) and $H\perp c$. For comparison, all spectra are normalized and shifted vertically. Solid lines in (a) are fits to the data using the linear frequency-field dependence of a paramagnet  [Eq.~(\ref{eq:EPR})], whereas solid lines in (b) denote simulations of resonance fields in the FM phase according to Eq.~(\ref{eq:omega_res}). Inset on panel (b) shows the low-frequency part of the $\nu(H_{\rm res})$ diagram on an enlarged scale.}
	\label{fig:f-dep}
\end{figure*}

The major distinction between ESR and FMR is that the former is the phenomenon of a resonant excitation of an ensemble of paramagnetic, possibly exchange coupled spins, whereas the latter is the resonance of the total magnetization of the ferromagnetically ordered sample. Since Cr$^{3+}$ (3$d^3$, $S =3/2$)  ions in \CGT\ have no orbital momentum in first order. Thus, their spectroscopic $g$ factor is generally expected to be very close to the spin-only value of 2 and only slightly anisotropic due to the second-order spin-orbit coupling effects \cite{AbragamBleaney} and interactions with the heavier Te-ligands, see discussion in Sec.~\ref{subsec:DFT}. Therefore, an ESR signal in the paramagnetic state of \CGT\ has to be almost isotropic with respect to the orientation of the applied external field.   In contrast, an FMR signal of \CGT\ in the ordered state may be appreciably anisotropic due to the magnetic shape anisotropy of the non-spherical sample as well as due to the intrinsic magnetocrystalline anisotropy, which both affect the resonance behavior of the magnetization of the sample depending on the particular direction of the magnetic field \cite{Turov,Skrotskii1966,Farle1998}. This expectation has been in fact confirmed by measurements of the frequency dependence of the position of the resonance signals for $H \parallel c$ and $H \perp c$ field geometries at two selected temperatures of 120\,K and 4\,K, i.e., well above and well below the FM transition temperature $T_{\rm c} \sim 66$\,K, respectively (Fig.~\ref{fig:f-dep}). 

In the paramagnetic state above $T_{\rm c}$ the resonance field $H_{\text{res}}$ shows a linear dependence on the microwave frequency $\nu$, as expected for a paramagnetic resonance:
\begin{equation}
\label{eq:EPR}
\nu = g\mu_B\mu_0H_{\text{res}}/h \ \ \ .
\end{equation}
Here $\mu_B$ denotes the Bohr magneton, $\mu_0$ is the vacuum permeability, $h$ is the Planck constant, and $g$ is the (effective) $g$-factor of the resonating spins. Indeed, only a small anisotropy between the two field orientations was observed at 120\,K and by fitting of Eq.~(\ref{eq:EPR}) to the data shown in Fig.~\ref{fig:f-dep}\,(a) the $g$-factors $g_{||} = 2.040 \pm 0.005$ and \mbox{$g_{\perp} = 2.023 \pm 0.005$} were obtained for field applied parallel and perpendicular to the $c$ axis, respectively (the respective fits are plotted by solid lines in Fig.~\ref{fig:f-dep}\,(a)). As will be discussed in the following in more detail, our ESR study reveals the existence of short-range FM correlations already above $T_{\rm c}$. In low-dimensional spin systems this can give rise to a frequency-dependent shift of the resonance line \cite{Okuda1972}. Thus, the $g$-factors determined from Eq.~(\ref{eq:EPR}) are, in fact, effective $g$-factors that contain both, the ionic $g$-factor of the magnetic ions and the additional contribution arising from low-dimensional spin correlations. 

At 4\,K, the frequency dependence exhibits a pronounced anisotropy as well as an orientation-dependent non-zero intercept with the ordinate of the frequency-field diagram, see Fig.~\ref{fig:f-dep}\,(b). Such a behavior is typical for FMR with magnetic fields applied parallel or perpendicular to the magnetic anisotropy axis \cite{Farle1998}. Thus, the $\nu(H_{\rm res}$) dependence of the FMR signal measured over a wide frequency range from 10 to 330\,GHz gives an additional proof for the identification of the $c$ axis as the magnetic easy axis, in line with the static magnetization measurements and previous observations in Refs.~\cite{Carteaux1995,Gong2017,Zhang2016}. Note that solid lines in Fig.~\ref{fig:f-dep}~(b) represent the simulation results using a standard phenomenological theory of FMR, as will be discussed in detail in Sec.~\ref{subsec:simulations}.

In order to gain detailed insights into the development of the magnetic anisotropy upon a transition from the high-temperature paramagnetic state into the low-temperature ferromagnetically ordered state, the shift $\delta H(T)$ of the resonance line from its paramagnetic resonance position was measured as a function of temperature at several selected excitation frequencies $\nu$ (Fig.~\ref{fig:T-dep}). We define the resonant shift as $\delta H(T) = H_{\rm res}(T) - H_{\rm res}$(120\,K), where the reference resonance field  $H_{\rm res}$(120\,K) was calculated according to Eq.~(\ref{eq:EPR}) using the $g$-factor values obtained from the frequency dependent measurements in the \mbox{paramagnetic state at $T = 120$\,K [Fig.~\ref{fig:f-dep}\,(a)].}
 
As can be seen in Fig.~\ref{fig:T-dep}, at all measured frequencies the signal is shifted to lower fields (\mbox{resulting} in a negative shift) if the magnetic field is oriented along the $c$ axis, which is the magnetic ``easy'' axis, while it is shifted to higher fields (corresponding to a positive shift) for magnetic field perpendicular to the $c$ axis. Despite the qualitative similarity between measurements at different frequencies concerning the observed trends of the shift $\delta H(T)$, the measured $\delta H(T)_{\parallel,\perp}$ dependences at distinct frequencies differ from each other regarding the details of the shape of the resulting $\delta H(T)$ curves. While the shift evolves rather smoothly for higher frequencies (open symbols in Fig.~\ref{fig:T-dep}\,(b)), measurements at 9.6\,GHz (open symbols in Fig.~\ref{fig:T-dep}\,(a)) revealed a much abrupter change of $\delta H(T)$ below 80\,K. A similar behavior was observed in the temperature dependent magnetization measurements $M(T)_{\parallel,\perp}$ at 0.35 and 3\,T (full symbols in Fig.~\ref{fig:T-dep}), respectively. Though in particular for strong fields where the $M(H)$ dependence at $T\ll T_{\rm c}$ is saturated (Fig.~\ref{fig:M_vs_H}),  the $M(T)_{\parallel}$ and $M(T)_{\perp}$  curves are rather symmetric in shape, as expected, the resonance shifts are not symmetric with respect to the abscissa as the functional form of the resonance condition in the ferromagnetic case depends on the orientation of external field relative to the magnetic anisotropy axis \cite{Turov,Farle1998}.

\begin{figure}[h]
	\includegraphics[width=0.9\columnwidth]{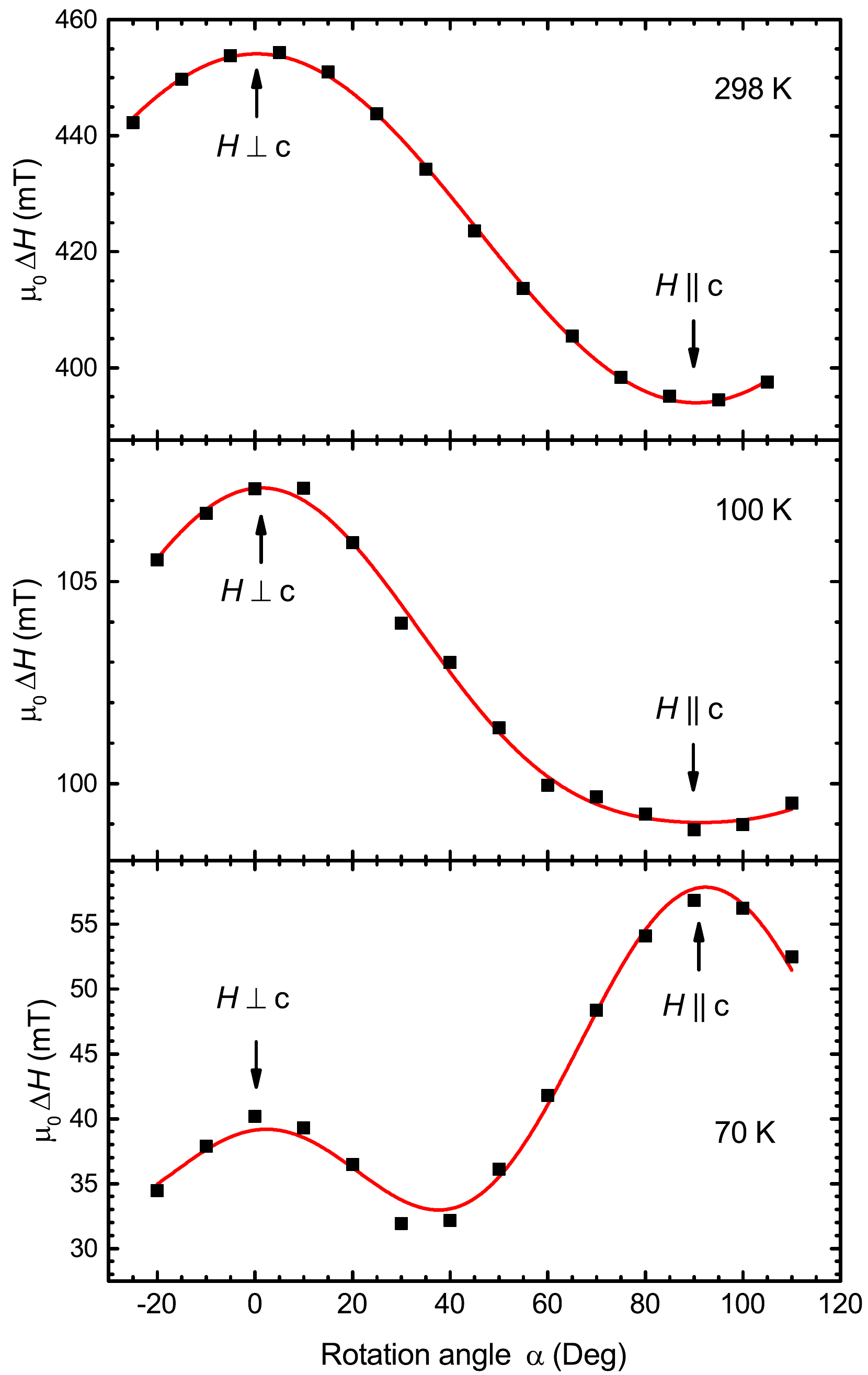}
	\caption{Angular dependence of the ESR linewidth obtained from measurements at 9.6\,GHz and temperatures of 298\,K (upper panel), 100\,K (center panel), and 70\,K (lower panel), respectively. Solid red lines are fits to the data as described in the main text.}
	\label{fig:ang_dep_X-band}
\end{figure}

It is worthwhile mentioning that the resonance lines are shifted from the paramagnetic position already below 100\,K, i.e. above $T_c$. In a classical ferromagnet, a temperature-dependent resonance shift is caused by internal magnetic fields resulting from an increasingly static net magnetization which, in turn, arises from FM couplings between the spins below the ordering temperature, see, e.g., Ref.~\cite{Skrotskii1966}. The fact that the $\delta H(T)$ dependence sets in at higher temperatures suggests that the FM correlations in \CGT\ on the typical ESR timescale of the order of $\nu^{-1}\sim 10$\,ps develop at temperatures significantly exceeding the $T_{\rm c}$. Such a behavior is reminiscent of one-dimensional spin systems where shifts of the resonance positions have been investigated experimentally and theoretically \cite{Okuda1972,Nagata1972,Oshima1976,Brockmann2012,Zeisner2017} and were attributed to the reduced dimensionality as well as anisotropic magnetic interactions. Consequently, an effective reduction of the dimensionality in the spin system could be the origin of the observed temperature-dependent resonance shift. Indeed, clear-cut evidence for the two-dimensional character of the spin correlations in \CGT \ is obtained from measurements of the angular dependence of the linewidth $\Delta H(\alpha)$. Fig.~\ref{fig:ang_dep_X-band} shows the angular dependent linewidth at three different temperatures together with fits to the data (solid red lines) as detailed in the following. At room temperature, upper panel in Fig.~\ref{fig:ang_dep_X-band}, $\Delta H(\alpha)$ reveals a $(\cos^2(\alpha)+1)$-dependence with $\alpha$ being the angle between external magnetic field and the $ab$ plane. This kind of angular dependence is typical for spin systems in an uncorrelated, i.e. paramagnetic, phase in which no significant interactions between the spins exist in any direction and, as a consequence, the reduced dimensionality does not affect the spin dynamics reflected in $\Delta H(\alpha)$ \cite{Richards1974,Benner1990}. Upon lowering the temperature, spin correlations start to grow, leading to a gradual change of the angular dependence. At 100\,K, center panel in Fig.~\ref{fig:ang_dep_X-band}, a slight deformation of the $(\cos^2(\alpha)+1)$-dependence is visible around $H \parallel c$ which can be described by a small additional contribution to $\Delta H(\alpha)$ of the form $(3\cos^2(\pi/2-\alpha)-1)^2$. Finally, at 70\,K, lower panel in Fig.~\ref{fig:ang_dep_X-band}, the angular dependence of the linewidth reveals a pure $(3\cos^2(\pi/2-\alpha)-1)^2$-behavior without $(\cos^2(\alpha)+1)$-contribution. Note, that the maximal linewidth occurs for the external field oriented perpendicular to the $ab$ plane, which requires in our definition of the rotation angle a shift of $\alpha$ by 90$^{\circ}$ in the argument of the trigonometric function. Such an angular dependence is typical for two-dimensional systems and is caused by the increasing dominance of long-wavelength fluctuations (or, in reciprocal space, $q \sim 0$ modes) in low-dimensional magnets, as discussed in Refs.~\cite{Richards1974,Benner1990,Chehab1991}. As at 70\,K no additional contributions to $\Delta H(\alpha)$ apart from the one related to the low dimensionality was observed, we conclude that the shift of the resonance line above $T_c$ is not caused by inhomogeneities in the sample, for which we also did not find any signature in the structural characterizations of the crystals, see Appendix~\ref{sec:sample_details}. Thus, a continuous slowing down of the correlated FM dynamics of the Cr spins by approaching the ordering transition from above, as evidenced by the onset of a resonance shift and a gradual change in $\Delta H(\alpha)$, demonstrates the intrinsically two-dimensional nature of the magnetism in \CGT, even in bulk crystals. Moreover, the two-dimensionality of magnetism as inferred from the analysis of static magnetization in the immediate vicinity of $T_{\rm c}$, see Refs.~\cite{Liu2017,Lin2017,Liu2018}, is found to be extended to much higher temperatures in the dynamic regime probed by ESR.

\begin{figure*}[!htbp]
	\centering
	\includegraphics[width=\textwidth]{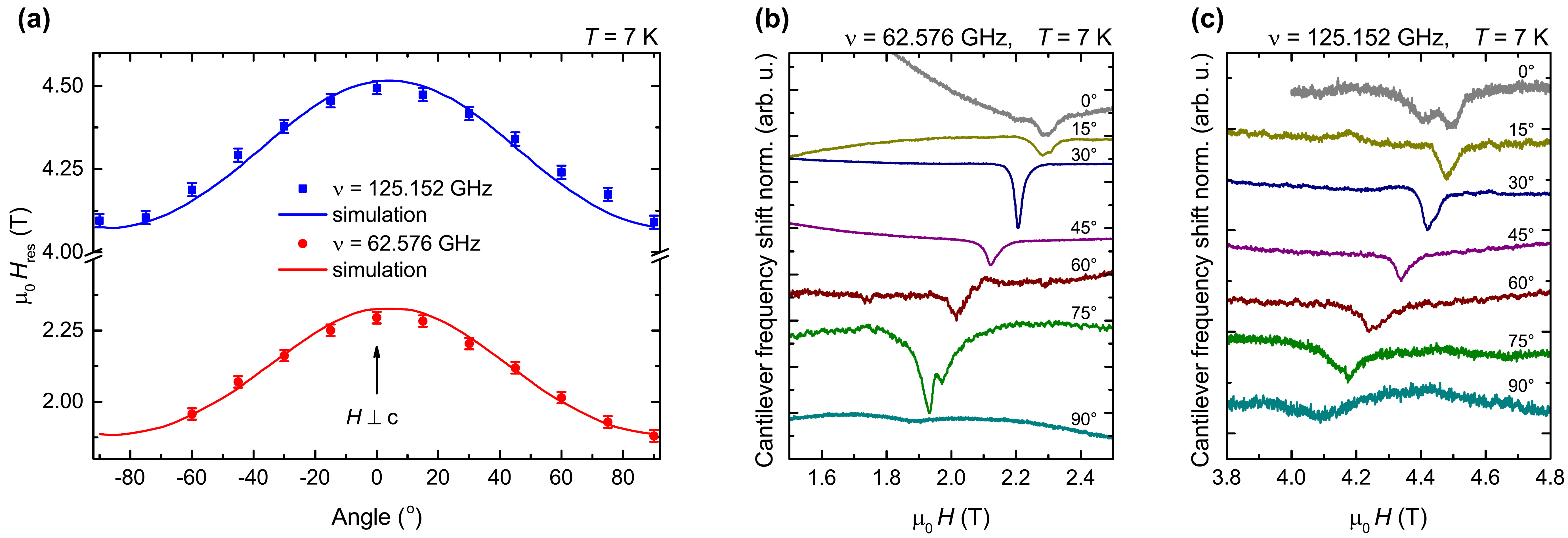}
	\caption{Angular dependence of the resonance field $H_{\rm res}(\theta)$ obtained from torque detected FMR at 7\,K and excitation frequencies $\nu$ of 62.6\,GHz and 125.2\,GHz (a). Solid lines are results of the simulations according to Eq.~(\ref{eq:omega_res}). Representative spectra obtained from measuring the shift of the cantilever frequency at $\nu = 62.6$ and 125.2\,GHz are shown in (b) and (c), respectively. For comparison, all spectra are normalized and shifted vertically.}
	\label{fig:ang-dep_canti}
\end{figure*}

In addition to the frequency dependent studies of  magnetic resonance signals in \CGT\ and the angular dependence of the linewidth at low frequencies above $T_c$ we investigated the angular dependence of the resonance field $H_{\rm res}(\theta)$ at 7\,K, i.e., deep in the FM phase, and at frequencies of 62.6\,GHz and 125.2\,GHz with a cantilever-based torque detected HF-ESR/HF-FMR setup. In this experiment the cantilever with the sample was rotated in the external magnetic field from $H \parallel c$ ($\theta = 90^\circ$) to $H \perp c$ ($\theta = 0^\circ$) directions. The obtained variations of the resonance position as a function of the angle $\theta$ between the vector of external magnetic field and the crystallographic $ab$ plane and exemplary spectra are shown in Fig.~\ref{fig:ang-dep_canti}. 

For both studied frequencies a similar sinusoidal behavior of $H_{\rm res}(\theta)$ was found: the maximal resonance field was observed when the magnetic field was aligned in the $ab$ plane, while a minimal resonance position corresponds to the magnetic field oriented parallel to the $c$ axis, which further supports a magnetic easy-axis scenario. Moreover, as the shift $\delta H$ of $H_{\rm res}$ from the paramagnetic value is governed by the magnetic anisotropy, angular dependent measurements allow a further independent determination of the anisotropy constants. Thus, the precision of the obtained value for the MAE could be increased by a combined evaluation of $H_{\rm res}(\nu)$ and $H_{\rm res}(\theta)$ data. For simulation of experimental data the same approach as for the frequency dependent measurements was employed, as described in Sec.~\ref{subsec:simulations}. Numerical results are shown as solid lines in Fig.~\ref{fig:ang-dep_canti}\,(a) together with the measured $H_{\rm res}(\theta)$ dependences. 

We note that the general qualitative behavior of the angular dependence of $H_{\rm res}$ is consistent with the one found in the low-frequency FMR/ESR experiments at $\nu = 9.6$\,GHz at temperatures between 120 and 50\,K (see Fig.~\ref{fig:shift_x-band}\,(a) in Appendix~\ref{sec:add_x-band}). Unfortunately, with further lowering the temperature the FMR signal at 9.6\,GHz corresponding to magnetic fields of about 0.35\,T became more and more structured and finally acquired a very distorted shape which prevents a reliable analysis of the data (see Fig.~\ref{fig:shift_x-band}\,(b) in Appendix~\ref{sec:add_x-band}). Most probably magnetic domains in the crystal are responsible for the observed distorted lineshapes at lower frequencies since magnetic fields required for full polarization of the magnetization in the sample are larger than the external field applied in this low-frequency magnetic resonance experiment (cf. Fig.~\ref{fig:M_vs_H}). In contrast, measurements at much higher frequencies $\nu$ corresponding to the fields above the saturation field yielded well defined single or, in the case of some cantilever measurements, double line spectra without any distortions down to the lowest measured temperature of 4\,K  (Figs.~\ref{fig:f-dep} and \ref{fig:ang-dep_canti}). Therefore, only HF-FMR data were used as the input for the simulations presented in the following section as the employed model is strictly applicable only for the case of the fully saturated magnetization.

\subsection{Simulations and determination of the magnetic anisotropies}
\label{subsec:simulations}

Simulation of the low-temperature frequency and angular dependences of the resonance field $H_{\rm res}$ [Fig.~\ref{fig:f-dep}\,(b) and Fig.~\ref{fig:ang-dep_canti}\,(a)] with a phenomenological model of ferromagnetic resonance yields reliable quantitative information about the magnetic anisotropies of \CGT. Within this well established approach (see, e.g., Refs.~\cite{Smit1955, Skrotskii1966, Farle1998}), the resonance frequency $\nu_{\text{res}}$ is given by the following equation:
\begin{equation}
\label{eq:omega_res}
\nu_{res}^2 = \frac{g^2 \mu_B^2}{h^2 M_s^2\sin^2\theta} \bigg(\frac{\partial^2{F}}{\partial \theta ^2}\frac{\partial^2{F}}{\partial \varphi^2} - \Big(\frac{\partial^2{F}}{\partial \theta \partial \varphi}\Big)^2\bigg) \ \ \  .
\end{equation}
Here, $g$ denotes the $g$-factor and $M_s$ represents the saturation magnetization of the sample, $F$ is a free energy density, and $\theta$ and $\varphi$ are the spherical coordinates of the magnetization vector $\boldsymbol{M}(M_s, \varphi, \theta)$. In order to calculate the resonance position for a given orientation of the sample in the external magnetic field and for a given microwave frequency Eq.~(\ref{eq:omega_res}) has to be evaluated at the equilibrium position $(\varphi_0, \theta_0)$ of $\boldsymbol{M}$. The latter is found via minimization of $F$ with respect to $\theta$ and $\varphi$ under the specific experimental conditions. In the present case, a complete description of our frequency and angular dependent HF-FMR data could be achieved taking into account a uniaxial magnetic anisotropy as well as shape anisotropies of the respective samples used for the measurements. The free energy density then reads
\begin{equation}
\label{eq:free_energy_density}
\begin{split}
F &= -\boldsymbol{H}\cdot\boldsymbol{M} - K_U\cos^2(\theta) + \\ & \ \ \ \ 2\pi M^2_s(N_x\sin^2(\theta)\cos^2(\phi) + \\ & \ \ \ \ N_y\sin^2(\theta)\sin^2(\phi) + N_z\cos^2(\theta)).
\end{split}
\end{equation}
Here, the first term describes the Zeeman energy density of the sample in the external field $\boldsymbol{H}$. The second term represents the uniaxial anisotropy characterized by the energy density $K_U$. The last term denotes the shape anisotropy with $N_x$, $N_y$, $N_z$ being demagnetization factors \cite{Osborn1945,Cronemeyer1991}. The specific values for the demagnetization factors of the two samples used for determination of $K_U$ are given in Table~\ref{tab:sample_dimensions} of Appendix~\ref{sec:sample_details} together with the sample dimensions.

As can be seen in Figs.~\ref{fig:f-dep}\,(b) and \ref{fig:ang-dep_canti}\,(a) it is possible to describe our HF-FMR results very well employing the model presented above.  It should be noted that the fields used in the HF-FMR experiments are larger than the saturation fields of the magnetization $H_{\rm sat}^{ab}$ and $H_{\rm sat}^{c}$ determined from the $M(H)$ curves in Sec.~\ref{subsec:magnetization}. This  justifies the use of this model which is applicable in the saturated regime of a ferromagnet. We emphasize that the simulated curves for all data sets were obtained using the same values for $K_U$ and $g$. For the latter, we used the $g$-factors determined from  HF-ESR data at 120\,K as starting values for the simulations (see Sec.~\ref{subsec:results_ESR}). These values were refined in order to achieve the best agreement between the simulated curves and the low temperature experimental data yielding at the end an isotropic $g$-factor of $2.040\pm0.005$, which is identical with the measured \mbox{$g_{||}$-factor}. 

For a complete description of all frequency and angular dependent HF-FMR data it was merely necessary to correctly account for the magnetic shape anisotropy which is determined by the particular geometry of the studied crystals. This was realized by calculating the proper demagnetization factors based on the measured dimensions of the crystals \cite{Osborn1945, Cronemeyer1991}, see Appendix~\ref{sec:sample_details}. As a result a value for $K_U$ of $(0.48 \pm 0.02) \times 10^6$\,erg/cm$^3$ was obtained which corresponds to an anisotropy energy of $(0.250 \pm 0.005)$\,meV per unit cell Cr$_6$Ge$_6$Te$_{18}$, comprising three formula units of \CGT. The value of $K_U$ is positive, which ultimately proves the magnetic easy axis anisotropy of \CGT, suggested by the magnetometry data, and other qualitative experimental observations. The magnetic anisotropy easy axis is collinear with the $c$ axis, i.e., it is directed perpendicular to \CGT\ layers. It is worth to mention that the obtained $K_U$ value is larger than the one previously reported in Ref.~\cite{Zhang2016}.  The difference is likely to be ascribed to the influence of the shape anisotropy which was not taken into account in Ref.~\cite{Zhang2016} and a saturation magnetization of about 2.23\,$\mu_B$/Cr reported by these authors. This value is lower than the theoretically expected one of 3\,$\mu_B$/Cr which has been confirmed by our measurements, see Sec.~\ref{subsec:magnetization}. Indeed, simulations of our data without shape-anisotropy term in Eq.~(\ref{eq:free_energy_density}) yield a $K_U$ of $(0.29\pm0.03)\times10^6$\,erg/cm$^3$ which is closer to the value reported in Ref.~\cite{Zhang2016} for low temperatures. However, for matching the simulation to the data it was necessary to adjust the $g$-factors for both samples individually. Moreover, the agreement between experimental data and simulations was worse as in the case of a proper treatment of the sample shape.
Therefore, it is crucial to consider the shape anisotropy in order to obtain a correct value for the intrinsic uniaxial MAE which is, in turn, essential for the understanding of the salient details of the magnetic properties of \CGT, in particular the origin of the easy axis type anisotropy. 

\subsection{DF calculations}
\label{subsec:DFT}
	
\begin{figure}
	\includegraphics[width=\columnwidth]{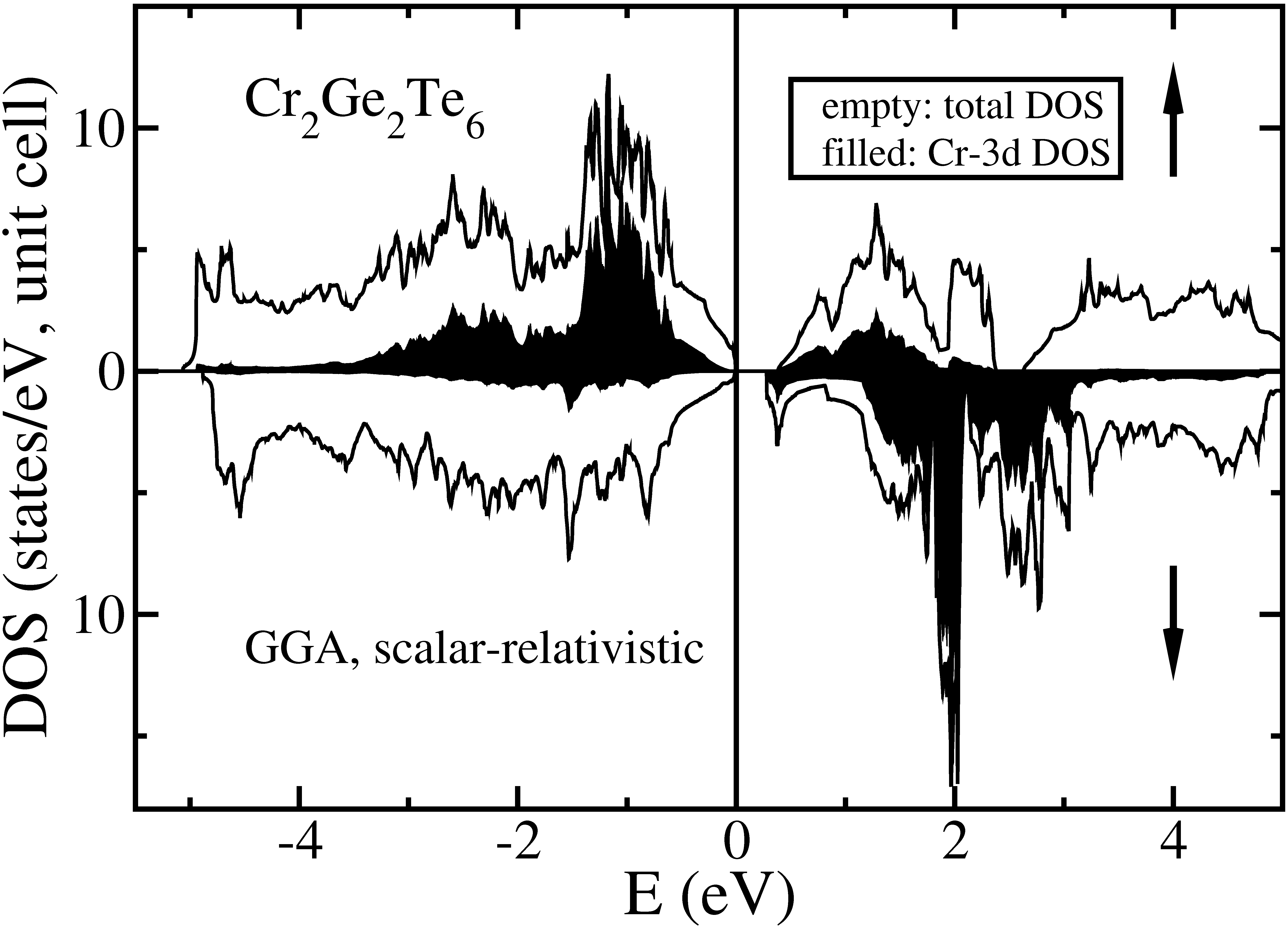}
	\caption{Density of states (DOS) of Cr$_2$Ge$_2$Te$_6$ obtained by
		a scalar-relativistic GGA calculation in a ferromagnetic state. 
		Arrows indicate the majority
		(up-arrow) and minority (down-arrow) spin channels. The lines limiting the
		empty areas denote the total spin-projected DOS and the filled areas denote
		the Cr-$3d$ spin-projected DOS.
	}
	\label{fig:scal_gga_bs}
\end{figure}

\begin{figure}
	\includegraphics[width=\columnwidth]{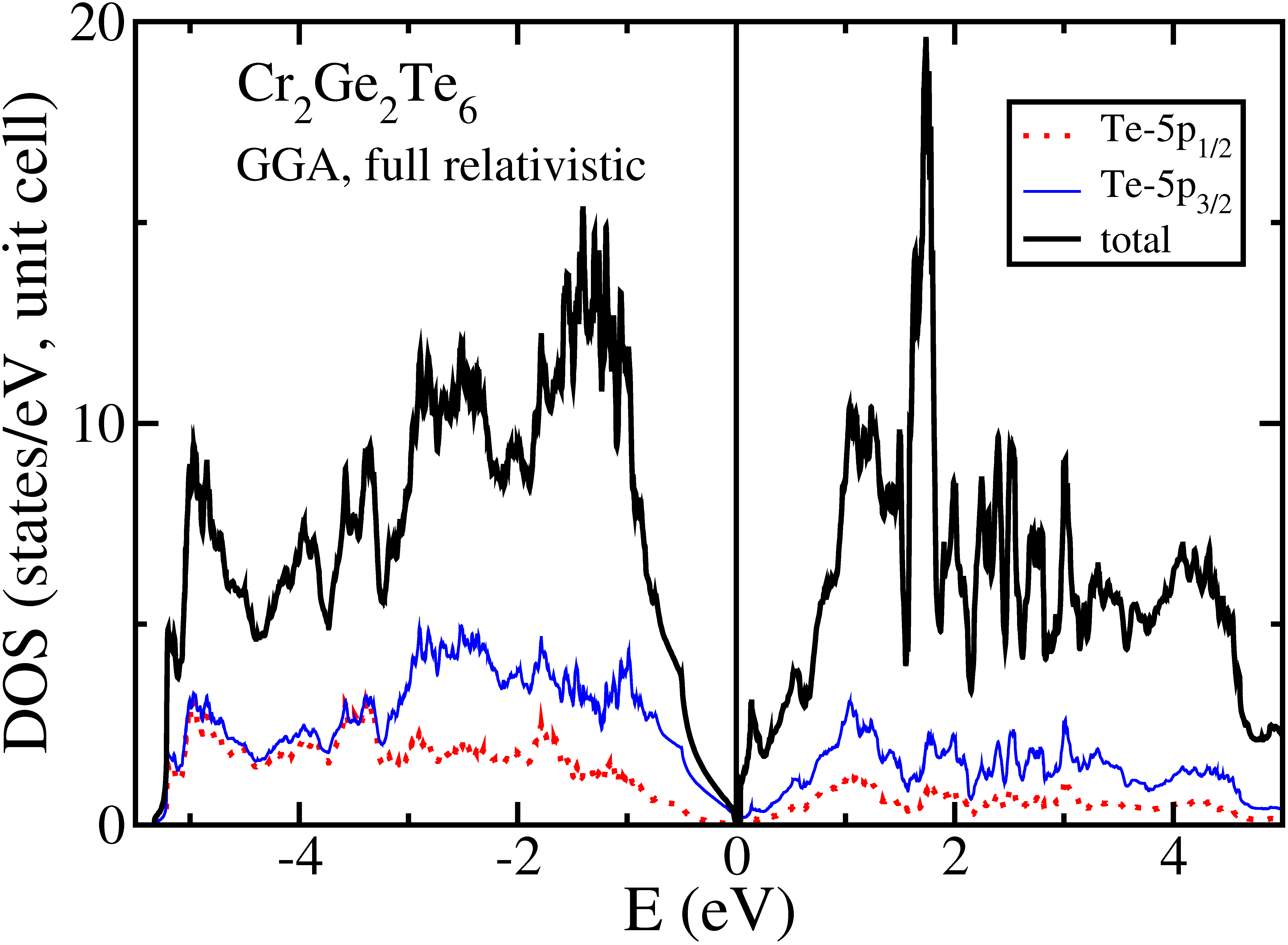}
	\caption{Density of states (DOS) of Cr$_2$Ge$_2$Te$_6$ obtained by
		a full relativistic GGA calculation carried out in a ferromagnetic state
		with moment orientation along $[100]$, i.e., within the easy plane. 
		Thick lines denote the total DOS,
		for both spin channels added up, thin lines denote the Te-$5p_{3/2}$
		projected DOS, and dotted lines denote the Te-$5p_{1/2}$ projected DOS.
	}
	\label{fig:GGA_fr_Te_p}
\end{figure}

The scalar-relativistic electronic structure of the title compound 
has been discussed in some detail already two decades ago in
Ref. \cite{Siberchicot1996}.
In particular, a gap of 0.06\,eV was found in the mentioned calculation,
separating the occupied majority-spin Cr-$3d$-$t_{2g}$ states from the other,
unoccupied Cr-$3d$ states and ensuring an integer total spin moment of 
3 $\mu_{\rm B}$ per Cr atom.
The present scalar-relativistic GGA results confirm this picture
of a ferromagnetic band insulator with a somewhat larger gap of 0.28\,eV,
Fig. \ref{fig:scal_gga_bs}.
However, if spin-orbit coupling is taken into account 
by switching to the full relativistic mode, the gap is almost closed,
Fig. \ref{fig:GGA_fr_Te_p}. The remaining narrow gap of 0.03\,eV is consistent with the experimental
transport gap of 0.007\,eV, measured below the ferromagnetic
ordering temperature~\cite{Carteaux1995}.

\begin{figure}
	\includegraphics[width=\columnwidth]{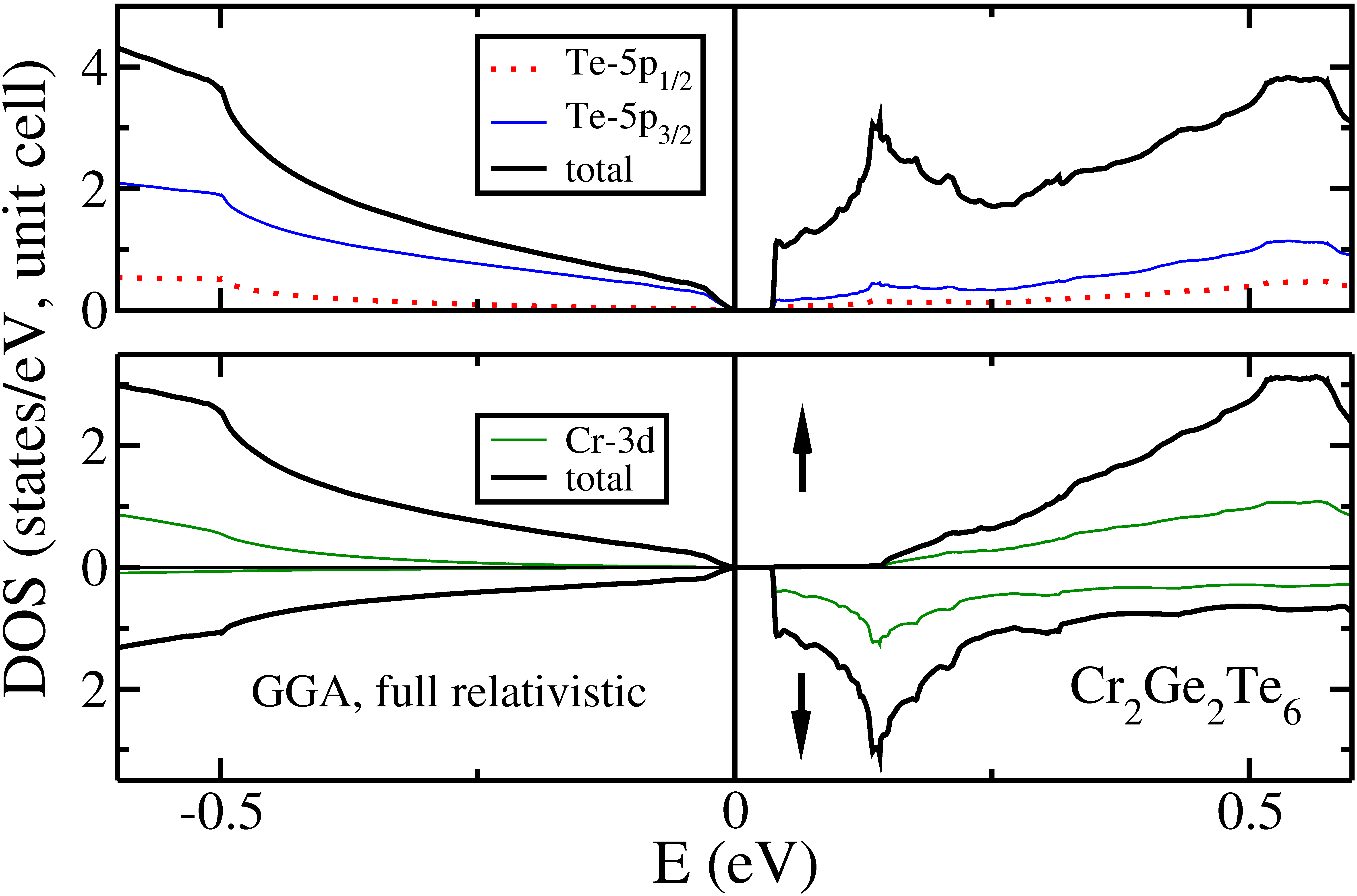}
	\caption{Density of states (DOS) of Cr$_2$Ge$_2$Te$_6$ obtained by
		a full relativistic GGA calculation carried out in a ferromagnetic state
		with moment orientation along $[100]$. Upper panel: same data as in
		Fig. \ref{fig:GGA_fr_Te_p} at an expanded energy-scale. Note the
		dominance of Te-$5p_{3/2}$ states at the top of the valence band and
		the asymmetric shape of the gap with a two-dimensional van-Hove singularity
		at the bottom of the conduction band. Lower panel: Spin-resolved presentation
		showing the total DOS (thick lines) and the Cr-$3d$ projected DOS.
		Note the prevalence of minority-spin states at the bottom of the conduction
		band.
	}
	\label{fig:GGA_fr_detail}
\end{figure}

The unusual reduction of the gap by spin-orbit coupling
can be understood by analyzing the orbital character of the states 
in the vicinity of the Fermi level, see Figs. 
\ref{fig:GGA_fr_Te_p} and \ref{fig:GGA_fr_detail}. 
The top of the valence band is dominated by Te-$5p_{3/2}$ states, 
while the bottom of the conduction band hosts a hybrid of Cr-$3d$, Te-$5p$, and
Ge-$4p$ states (the latter are not shown) with contributions 
to the total density of states of 36\%, 21\%, and 27\%, respectively.
Being the heaviest element of the investigated compound, 
Te experiences the largest spin-orbit coupling. 
Its $5p$ states are split by about 0.8\,eV, both in the free atom and in
the pseudo-atom used for the construction of the valence basis for the title
compound.
Due to this splitting, the center of gravity of the Te-$5p_{3/2}$ states
moves upward in energy by about 0.2\,eV against the other atomic states, compared with the case without spin-orbit coupling.
In the bulk compound, this is visible by a related shift of band weights,
see Fig. \ref{fig:GGA_fr_Te_p}: Below -3\,eV, Te-$5p_{1/2}$ and Te-$5p_{3/2}$
states contribute almost the same weight to the DOS, while Te-$5p_{3/2}$ states
clearly dominate at higher energies including the region close to the gap.

\begin{figure}
	\includegraphics[width=\columnwidth]{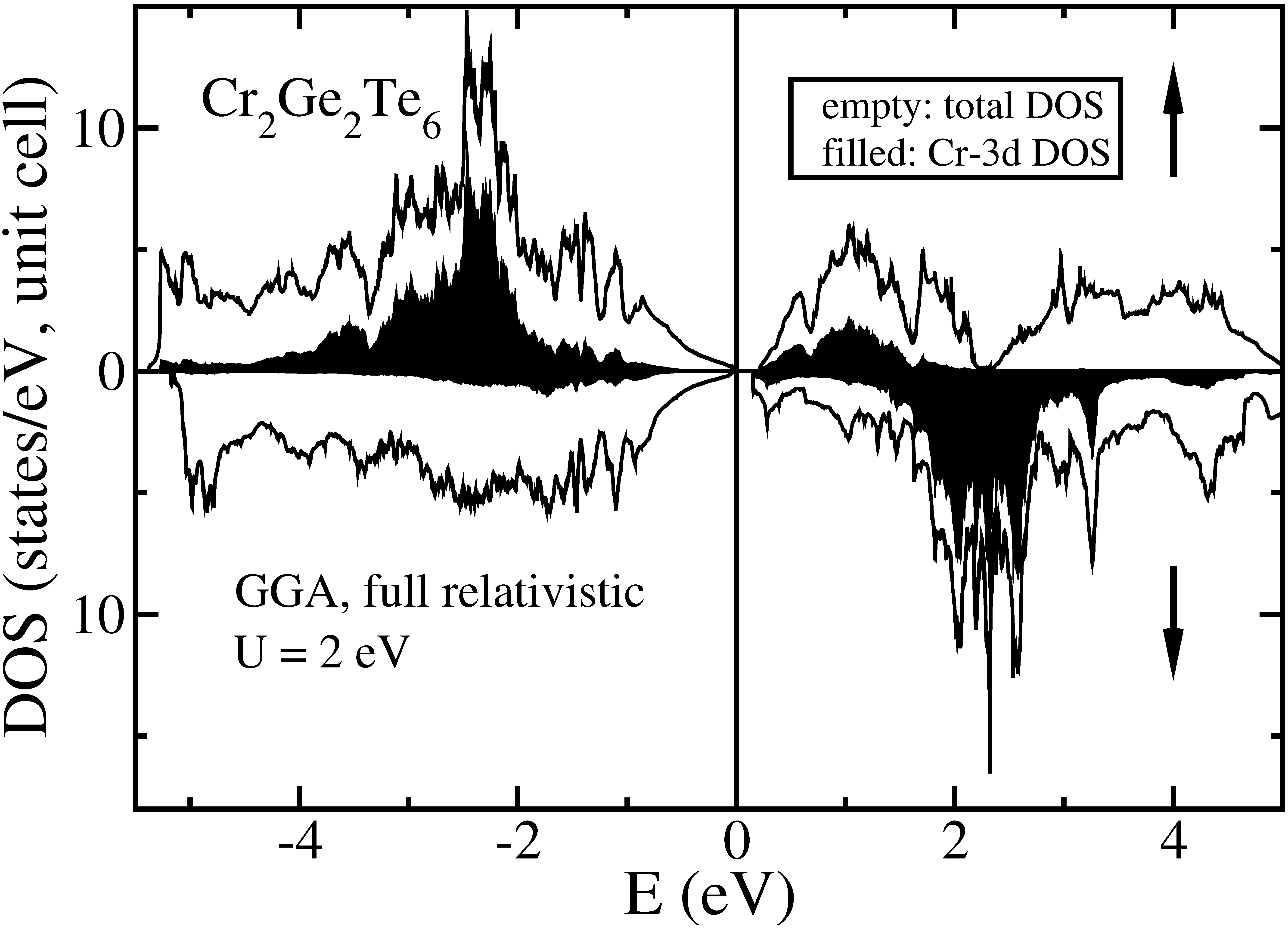}
	\caption{Density of states (DOS) of Cr$_2$Ge$_2$Te$_6$ obtained by
		a full relativistic GGA$+U$ calculation carried out in a ferromagnetic state
		with moment orientation along $[001]$, i.e., along the easy axis.
		The lines limiting the
		empty areas denote the total spin-projected DOS and the filled areas denote
		the Cr-$3d$ spin-projected DOS.
	}
	\label{fig:u2_gga_bs}
\end{figure}

\begin{figure}
	\includegraphics[width=0.9\columnwidth]{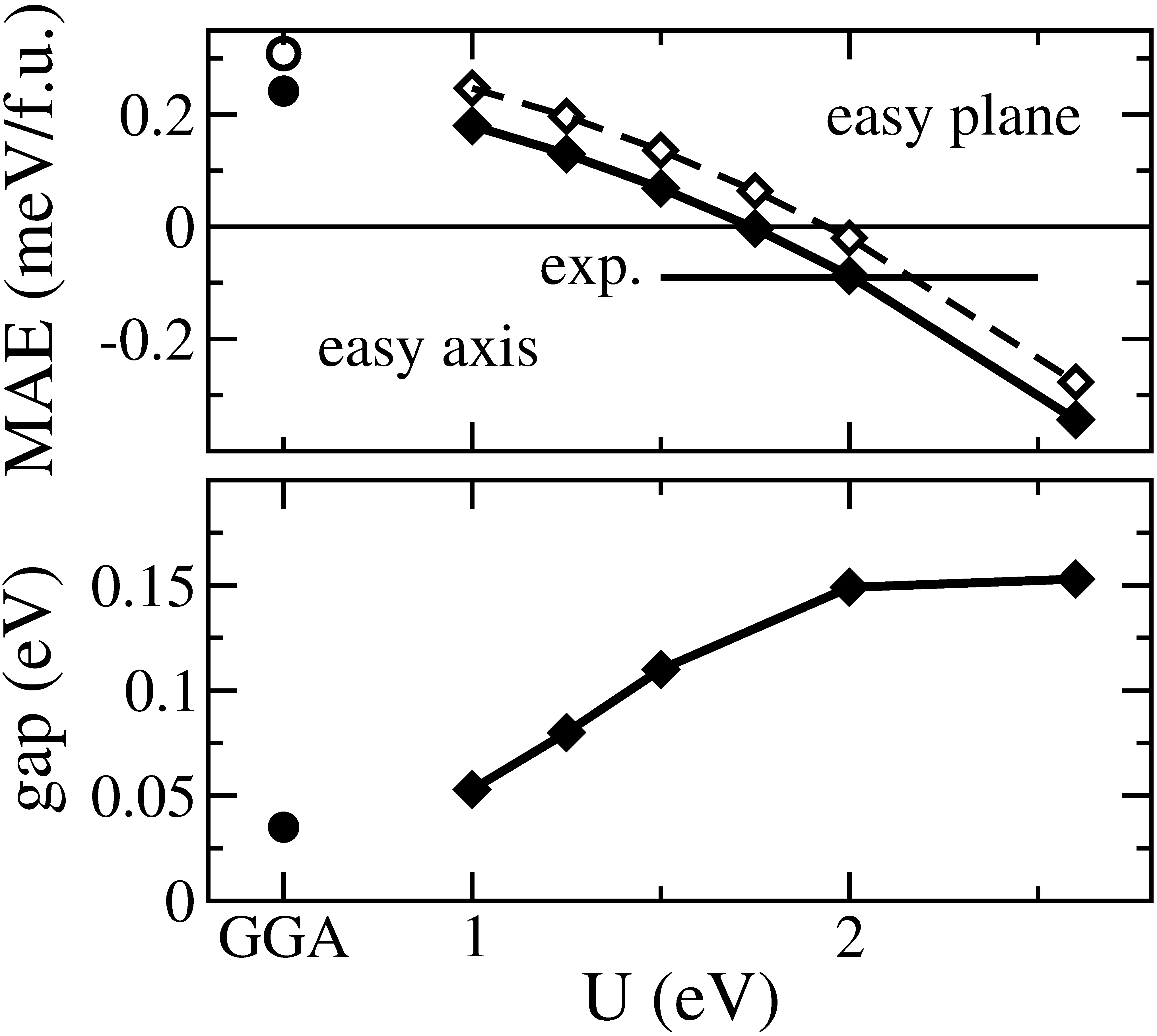}
	\caption{Upper panel: Calculated MAE of 
		Cr$_2$Ge$_2$Te$_6$ as obtained
		by GGA (full circle) and GGA+$U$ approaches for different values of $U$
		(diamonds connected by guiding lines). The MAE depends on $U$ monotonously for all considered values up to $U$ = 5\,eV. A horizontal line denotes the
		experimental value obtained in this work. Data with open symbols denote the total magnetic anisotropy energy, i.e., the sum of calculated MAE of this work and intrinsic dipolar anisotropy energy of 0.067\,meV/f.u. taken from Ref.~\cite{Fang2018}. Note, that the sign definition of MAE in Ref.~\cite{Fang2018} differs from our definition. Lower panel: Calculated gap size (same meaning of the symbols as in the upper graph). Note, $J$ = 0.6\,eV in all GGA+$U$ calculations. Thus, the GGA result is not the same as the limit $U \rightarrow 0$ of the GGA+$U$ data. The position of the GGA results on the abscissa is arbitrary.
	}
	\label{fig:gap_mae}
\end{figure}

An interesting peculiarity is observed at the bottom of the conduction
band, which is of almost pure (99\%) minority-spin character, 
see Fig. \ref{fig:GGA_fr_detail}. Essential majority spin population
sets in only about 100\,meV above the conduction band minimum.
The strong spin polarization at CBM
is in qualitative agreement with results of earlier GGA+$U$
calculations taking into account van der Waals interactions~\cite{Li2014}
and also with similar recent calculations for a few-layer \CGT \ 
system~\cite{Wang2018}. 
Different from those scalar-relativistic 
calculations we find that the top of the valence band is strongly
spin-mixed by the strong spin-orbit interaction within the Te-$5p$ shell
and that the CBM hosts minority-spin bands.

We note that the steep van-Hove singularity 
at the conduction band bottom indicates a quasi two-dimensional band.
As mentioned above, this band
is composed of orbitals from all constituents of the system. Thus, it can
be considered to host quasi-homogeneous states.
Occupation of these states with charge carriers will result in an almost
completely spin-polarized, quasi two-dimensional, and
quasi-homogeneous electron gas.

While the narrow experimental transport gap is reasonably well described 
by the full relativistic GGA data, the magnetocrystalline anisotropy (MA) 
turns out to be of easy-plane type in this approach, at variance with 
the easy-axis MA found in experiment, see Fig. \ref{fig:gap_mae}.
For this reason, further calculations were performed relying on the GGA+$U$ 
method. Depending on the specific value of $U$, the occupied and the unoccupied
Cr-$3d$ bands are shifted further away from each other, compare
Figs. \ref{fig:scal_gga_bs} and \ref{fig:u2_gga_bs}. 
This changes the relative positions
of spin-polarized Cr-$3d$ states and spin-orbit split Te-$5p$ states.
As a result, the MA strongly depends on the chosen value of $U$.
At about $U \approx$ 2\,eV, a reasonable value for Cr-$3d$ states in a narrow-gap system, the calculated MAE meets the measured value.
However, the calculated gap of about 0.15\,eV at $U$ = 2\,eV is more than an
order of magnitude larger than the experimental transport gap.

A similarly strong dependence $\delta$ MAE$/\delta U$ of about \mbox{$3 \times 10^{-4}$/f.u.} (see Fig.~\ref{fig:gap_mae}) was reported for the title compound by C. Gong \textit{et al.} recently (cf. Extended Data Figure~8 of Ref.~\cite{Gong2017}). Those authors found a transition from easy-plane to easy-axis MA at $U$ = 0.2\,eV, to be compared with our value of 1.75\,eV. This offset can be understood by the fact that the authors of Ref.~\cite{Gong2017} used in their GGA+$U$ calculation only the Slater integral \mbox{$F^0$ = $U$}, while we included two further Slater integrals via \mbox{$J$ = ($F^2$ + $F^4$)/14} which, first, tends to reduce the effect of $U$ and, second, introduces anisotropic contributions to the correlation.

An additional contribution to the intrinsic magnetic anisotropy energy is due to the magnetic dipole interaction. 
The latter is known to be responsible for the magnetic shape anisotropy. Less known is that it also accounts for an intrinsic (bulk) contribution \cite{Jansen1988} which is usually very small \cite{Daalderop1990}. In layered materials, it may however provide larger contributions as has been demonstrated for the title compound recently \cite{Fang2018}. We included the additional term in Fig. 10 as a constant shift, as it only depends on the size of the atomic moments and on the structure, both being not very sensitive to the value of $U$.

It should be noted, that the polarized quasi two-dimensional band at
the conduction band bottom is robust with respect to changing the direction 
of magnetization and also with respect to the application of $U$ up to 2\,eV. 
For this value of $U$ and easy-axis orientation
of the magnetic moment, its distance from the lowest majority-spin conduction
band is reduced to
60\,meV as compared to 100\,meV in the GGA calculation with magnetic moment 
oriented along $[100]$. If $U$ is increased beyond 2\,eV, the minority Cr-$3d$ states are shifted further toward higher energies. At $U \approx 2.3$\,eV, the character of CBM switches to majority spin. While the CBM is still almost completely spin polarized (with the exception of the switching value of $U$), it is no longer quasi-2D.

\section{Conclusions}
\label{sec:conclusions}

In conclusion, the spin dynamics and magnetic anisotropy of the bulk form of the 2D van der Waals ferromagnet \CGT \ was investigated in detail by means of high-field ESR and FMR spectroscopies, magnetization measurements, and density functional calculations. Magnetic resonance data evidence a gradual onset of the ferromagnetic correlations at temperatures well above the magnetic phase transition which continuously grow and slow down by approaching the $T_{\rm c}$. Such a pronounced correlated spin dynamics appears to be a characteristic feature of the two-dimensional magnetism of \CGT. Both static and dynamic experimental methods directly evidence the easy-axis-type nature of the total magnetic anisotropy, including magnetocrystalline and shape anisotropy, in line with previous studies\cite{Carteaux1995,Zhang2016,Lin2017,Liu2017}. Simulations of the HF-FMR experiments as a function of microwave frequency and angle of the magnetic field with respect to the sample yield a MAE $K_U$ of $(0.48\pm0.02)\times10^6$\,erg/cm$^3$, explicitly taking into account the shape of individual crystals. Thus, the obtained value represents an accurate magnitude of the intrinsic magnetocrystalline anisotropy energy density as compared to the value reported in  Ref.~\cite{Zhang2016} where the shape anisotropy apparently was not taken into account. Comparing the MAE obtained from phenomenological simulations of experimental data with DF calculations employing the GGA+$U$ method, we find a good agreement with the theoretical MAE for $U$ = 2\,eV. However, the gap calculated at this particular $U$ value differs from experimental observation. Analysis of the electronic structure indicates the presence of quasi two-dimensional states at the bottom of the conduction band. Upon electron doping~\cite{Wang2018}, these states may host a quasi-homogeneous, almost completely spin-polarized electron gas.

\section*{Acknowledgments}
This research was supported by the Deutsche Forschungsgemeinschaft (DFG) via grants \mbox{AL 1771/4-1} and \mbox{KA1694/8-1}. S.A. and S.S. acknowledge financial support from GRK-1621 graduate academy of the DFG. M.P.G. thanks the Alexander von Humboldt Foundation for financial support through the Georg Forster Research Fellowship Program. M.P.G. and M.R. are grateful for technical assistance by Ulrike Nitzsche.

\appendix

\section{Details on samples used in this study}
\label{sec:sample_details}

\begin{figure*}[!t]
	\includegraphics[width=0.8\textwidth]{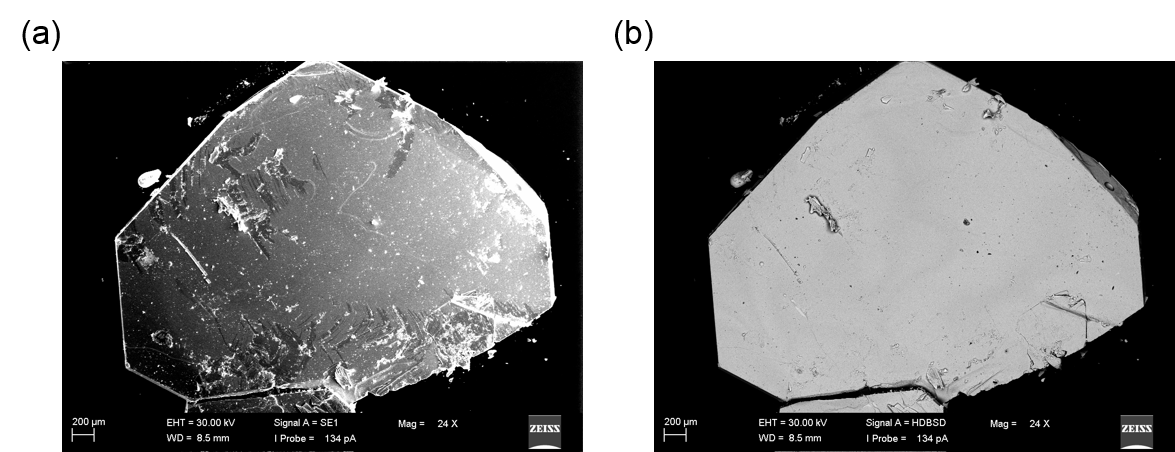}
	\caption{SEM images of a \CGT\ single crystal using the height contrast SE mode (a) and the chemical contrast BSE mode (b).}
	\label{fig:sample_BSE}
\end{figure*}

Detailed information about structural characterization of the synthesized \CGT \ crystals by means of PXRD at room temperature are given in Tab.~\ref{tab:Rietveld1}. The homogeneity of the crystals was confirmed by scanning electron microscopy (SEM) imaging using a high definition back scattered electron detector (HDBSD). SEM images using a secondary electron (SE) detector and a back scattered electron (BSE) detector are shown in Fig.~\ref{fig:sample_BSE}. Using a SE detector, the contrast in the image is given by differences in height and topology while using a BSE detector results in a contrast given by differences in chemical composition. The homogeneous color and lack of contrast of the SEM(BSE) image (Fig.~\ref{fig:sample_BSE}(b)) over a wide area indicates a homogeneous chemical composition of the sample and its purity. Nevertheless, some spots of different contrast can also be seen on the sample surface. If compared to the SEM(SE) image (Fig.~\ref{fig:sample_BSE}(a)), those spots are also showing a height contrast. They originate from dust and dirt particles on the sample surface and are not obtained due to a secondary phase in the sample.

\begin{table}[!b]
	\caption{Details of structural parameters and R-values of Rietveld analysis.}
	\setlength\extrarowheight{1.2pt}
	\begin{ruledtabular}
		\begin{tabular}{l l}
			Parameter & \CGT \\
			\hline
			Wavelength (\r{A}) & 1.54059  \\
			$\theta$ Range ($^\circ$) & 10.00\,-\,101.99 \\
			Step Size ($^\circ$) & 0.01 \\
			Temperature (K) & 293 \\
			Space Group & \textit{R}\={3} H (No. 148) \\
			\textit{a} (\r{A}) & 6.828(3) \\
			\textit{c} (\r{A}) & 20.572(12) \\
			U$_{isotropic}$: & \\
			U$_{Cr}$ & 0.041(9) \\
			U$_{Ge}$ & 0.016(6) \\
			U$_{Te}$ & 0.029(2) \\
			R$_p$ & 0.0476 \\
			R$_{wp}$ & 0.0615 \\
			R$_f$ & 0.0902 \\
			Goodness-Of-Fit & 1.24 \\
		\end{tabular}
	\end{ruledtabular}
	\label{tab:Rietveld1}
\end{table}

\begin{figure}[!bh]
	\centering
	\includegraphics[width=\columnwidth]{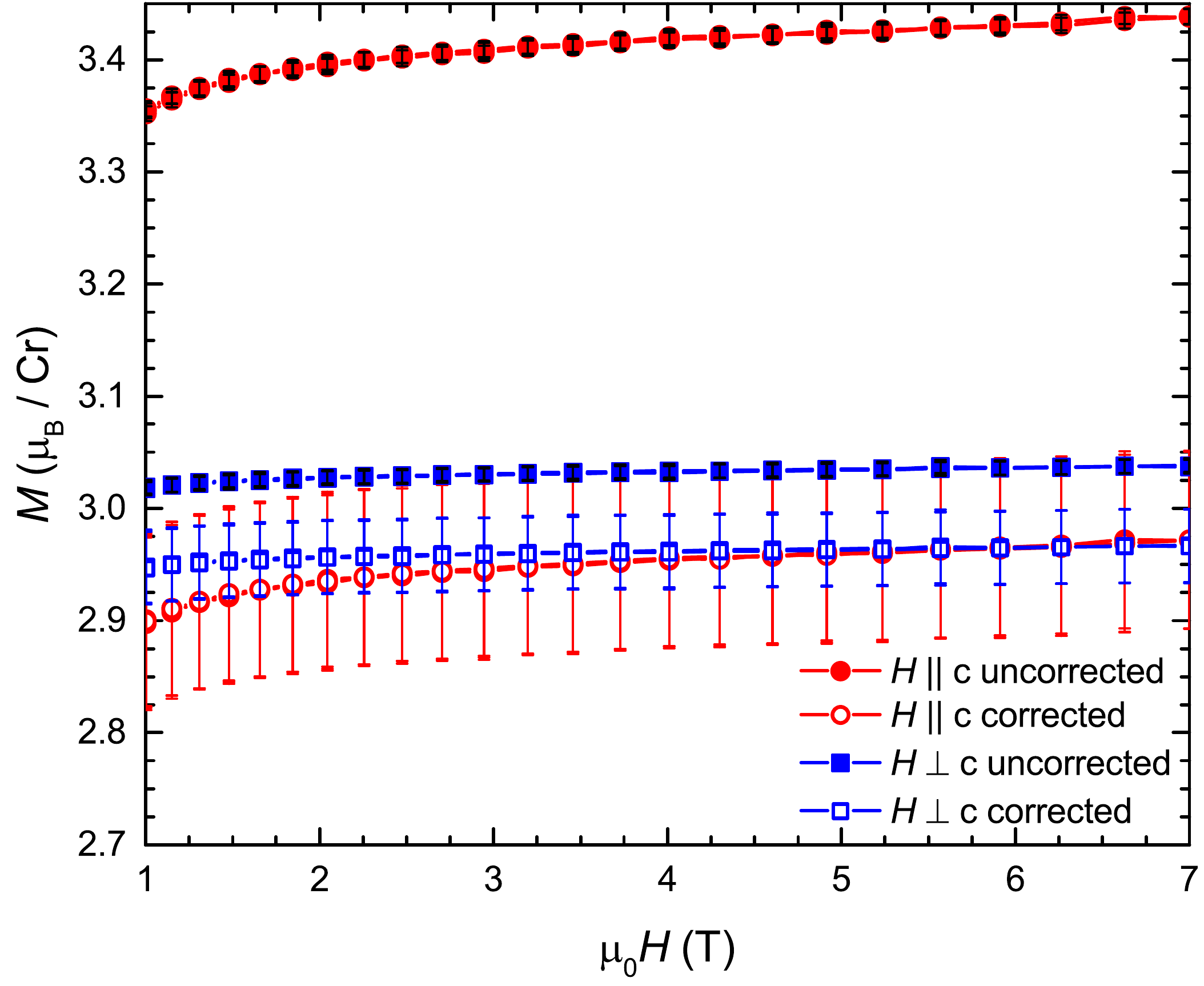}
	\caption{Example for the effect of magnetization correction for a field-dependent measurement of the magnetization at 4\,K for $H \parallel c$ (red circles) and $H \perp c$ (blue squares), respectively. The as-measured data are shown as full symbols, while open symbols represent the corrected data.}
	\label{fig:example_correction}
\end{figure}

DC magnetization measurements were carried out on a 4\,mm $\times$ 3\,mm $\times$ 2\,$\mu$m crystal. For the correction of the sample magnetic moment an equivalent surface area square film sample with a side length of 3.3\,mm and thickness of 2\,$\mu$m was assumed. The correction factors for this sample were obtained based on the experimental data on a squared nickel film presented in Ref.~\cite{VSM_corr}. To extract correction factors for the assumed side length of 3.3\,mm the data on the squared nickel film for the same vibration amplitude (2\,mm) was fitted by exponential functions yielding a good agreement between fit and experimental data (for $H$ $\parallel$ c (horizontal): $\chi^2$\,=\,7.7$\cdot$10$^{-6}$; for $H$ $\perp$ c (vertical): $\chi^2$\,=\,8.0$\cdot$10$^{-7}$). From these functions the factors of 1.157 ($\pm$0.031) for $H$ $\parallel$ c and 1.024 ($\pm$0.011) for $H$ $\perp$ c for the overestimation of the measured moment by the experimental setup were extracted. The measured moments were then divided by those factors to account for the non point-like dipole sample shape. As an example for the effect of this correction, Fig.~\ref{fig:example_correction} shows the uncorrected and corrected magnetization as function of field at 4\,K close to and in the saturation magnetization regime of \CGT. It is clearly visible that without this correction two different values of saturation magnetization for the different field directions are obtained. After correction, the difference in saturation magnetization for the different field directions vanishes. The strongly different values of saturation magnetization for the uncorrected case would indicate either a strongly anisotropic $g$-factor or an unusual interplay with a secondary magnetic phase. We can rule out both scenarios for the \CGT \ sample by our ESR and structural (PXRD and SEM-EDX) investigations. In fact, the isotropic saturation magnetization after the correction is in good agreement with the nearly isotropic $g$-factors obtained by ESR. Furthermore, the correction leads to a saturation moment close to 3\,$\mu_B$/Cr, which is the expected moment for Cr$^{3+}$ with $S$\,=\,3/2 and $g$\,$\sim$\,2 and is in good agreement with literature \cite{Carteaux1995,Ji2013,Lin2017,Liu2018}. We note that the use of the correction factors yield the theoretically expected results which can be understood as a positive self-consistency check of the applied correction method.

\begin{figure}[!bt]
	\centering
	\includegraphics[width=0.6\columnwidth]{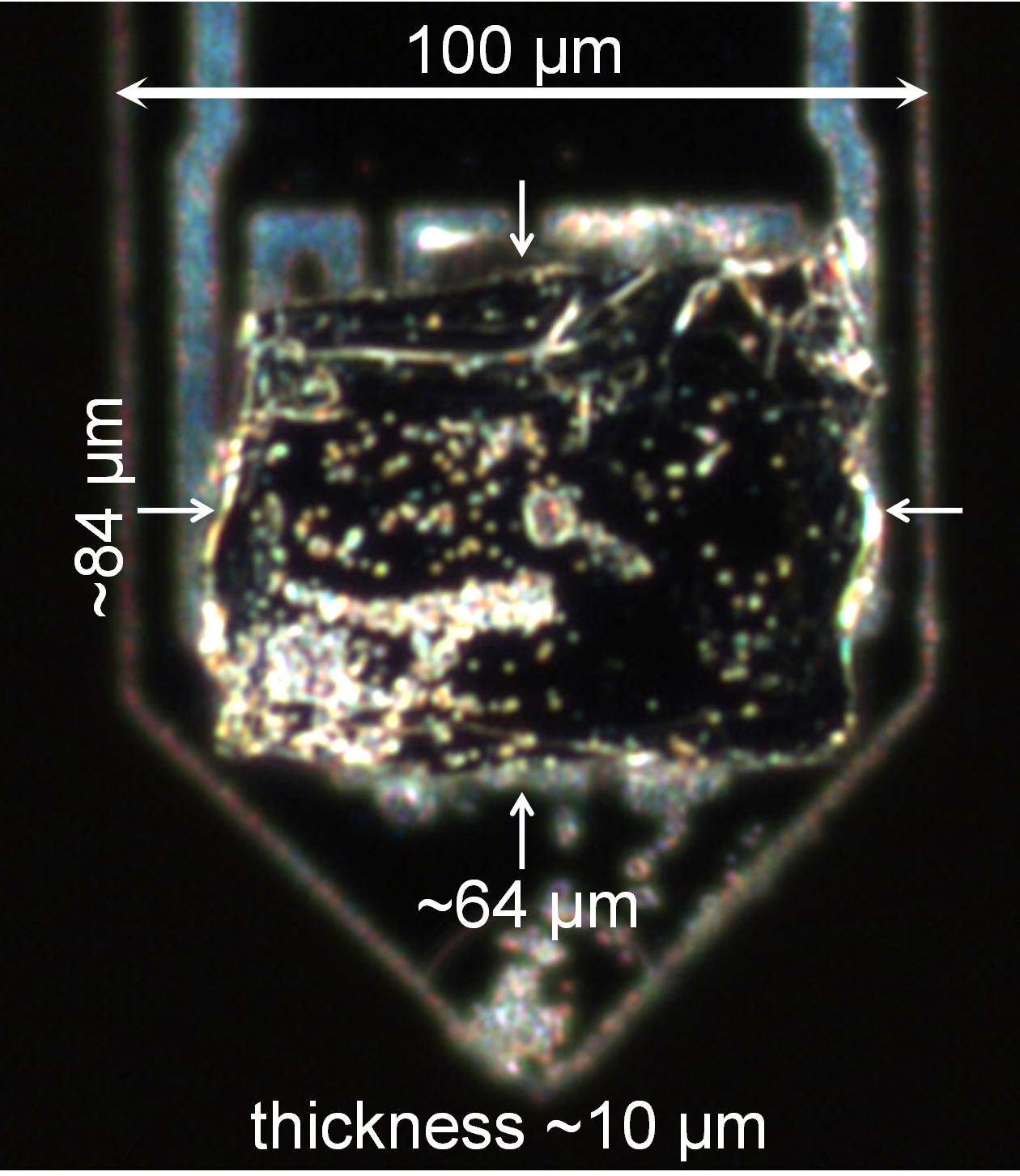}
	\caption{\CGT \ sample used for the torque detected FMR study presented in this work as mounted on a piezocantilever.}
	\label{fig:sampleoncanti}
\end{figure}

Cantilever-based FMR measurements were carried out on a small crystal (CL sample) mounted on the cantilever as shown in Fig.~\ref{fig:sampleoncanti}. Compared to this sample, the crystal employed in the main part of the X-band (9.6\,GHz) and HF-ESR measurements (HF sample) was much larger and was chosen in order to increase the signal intensity (note that the ESR intensity is proportional to the number of spins in the sample). For an evaluation of demagnetization factors, lengths of the sample edges were measured from which the dimensions $l$, $w$, and $h$ of an equivalent cuboid with volume $l \times w \times h$ were determined. The resulting dimensions are given in Table~\ref{tab:sample_dimensions} for both samples together with the respective demagnetization factors calculated based on the equivalent-ellipsoid method \cite{Cronemeyer1991,Osborn1945}. Uncertainty in measurements of the sample dimensions caused uncertainties in the calculated demagnetization factors which did not exceed 9\%. Note that deviations of the real sample shape from the approximated cuboid are small compared to the differences between the lateral dimensions $l$ and $w$ and the sample thickness $h$. 

\begin{table}[!tbh]
	\caption{Dimensions and demagnetization factors for the samples used in the FMR study.}
	\setlength\extrarowheight{0.5pt}
	\begin{ruledtabular}
		\begin{tabular}{c c c c c c c}
			& $l$ ($\mu$m) & $w$ ($\mu$m) & $h$ ($\mu$m) & $N_x$ & $N_y$ & $N_z$ \\
			CL sample & 84 & 64 & 10 & 0.1643 & 0.0523 & 0.7834\\
			HF sample & 1250 & 1030 & 35 & 0.0430 & 0.0017 & 0.9553\\
		\end{tabular}
	\end{ruledtabular}
	\label{tab:sample_dimensions}
\end{table}

\section{Additional data obtained from X-band ESR measurements}
\label{sec:add_x-band}

\begin{figure}[!tb]
	\centering
	\includegraphics[width=\columnwidth]{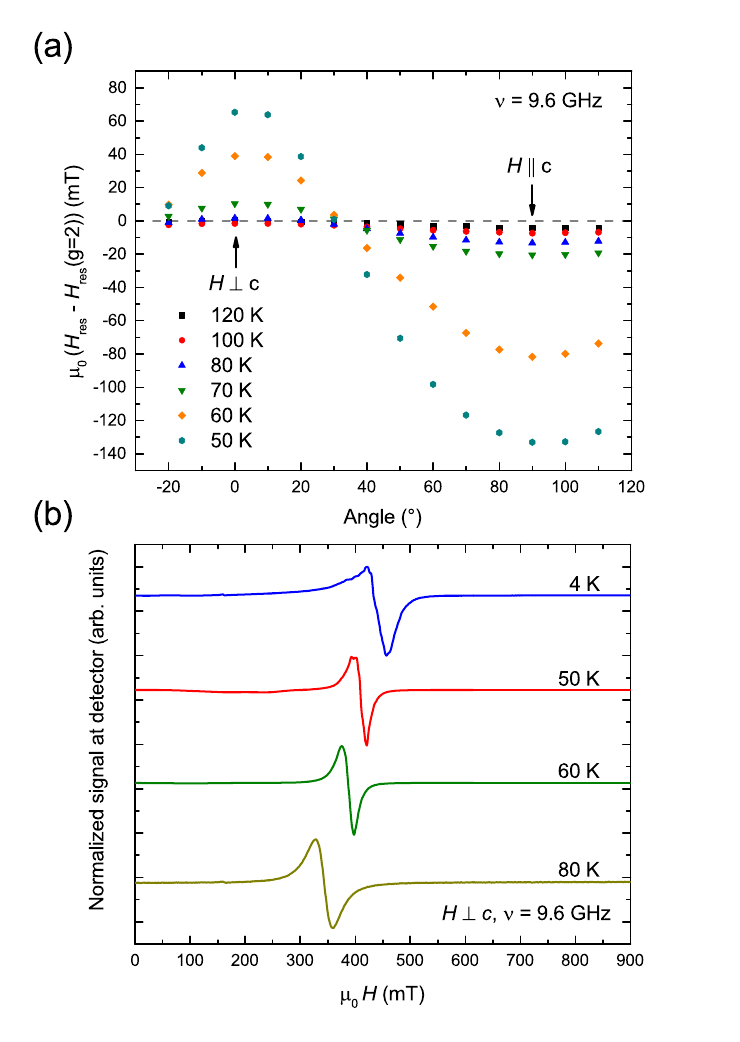}
	\caption{Angular dependence of the resonance shift $\delta H(\theta)$ (a) and representative spectra (b) at various temperatures and at a microwave frequency of 9.6\,GHz. The spectra shown in panel (b) are field derivatives of the respective absorption lines due to the use of the lock-in technique in this type of spectrometer. They were recorded with the external magnetic field applied perpendicular to the $c$ axis. For comparison, spectra are normalized and shifted vertically.}
	\label{fig:shift_x-band}
\end{figure}

In addition to the experimental data presented in Sec.~\ref{subsec:results_ESR} of the main text angular dependence of the resonance field $H_{\rm res}(\theta)$ was studied at a microwave frequency of 9.6\,GHz at several selected temperatures. Resonance shifts $\delta H$ obtained from these measurements are summarized in Fig.~\ref{fig:shift_x-band}\,(a). At 120\,K only a weak anisotropy, as reflected in the angular dependence of the resonance shift, was observed. Upon decrease of temperature the anisotropy increases due to the growth of FM correlations. Below 50\,K such an analysis of the anisotropy is hampered by a strong distortion of the line, as is illustrated by exemplary spectra in Fig.~\ref{fig:shift_x-band}\,(b). As discussed in the main text, such a distortion is most probably related to magnetic domains likely to be still present at small fields used in this experiment. Consequently, the resonance shift cannot be determined unambiguously at low frequencies below 50\,K.

\clearpage

\bibliography{literature_Cr2Ge2Te6}

\end{document}